\documentclass{aa}

\usepackage{txfonts}

\usepackage[english]{babel}
\usepackage{amsmath,amssymb}
\usepackage{graphicx}
\usepackage[normalem]{ulem}

\usepackage{verbatim}

\usepackage{multirow}
\usepackage{mathtools}
\usepackage{epstopdf}

\usepackage{url}
\usepackage{xcolor,color}

\usepackage{natbib}

\usepackage{tabularx}

\bibpunct{(}{)}{;}{a}{}{,} 
\definecolor{cobalt}{rgb}{0.06, 0.2, 0.65}
\usepackage[colorlinks=true,citecolor=cobalt,urlcolor=cobalt]{hyperref}
\hypersetup{
  colorlinks,
  citecolor=cobalt,
  linkcolor=[rgb]{0.8, 0.2, 1.0},
  urlcolor=cobalt,
}
\usepackage{orcidlink}

\usepackage{siunitx}
\usepackage{capt-of}
\usepackage{booktabs}
\usepackage{array}
\newcolumntype{C}[1]{>{\centering\arraybackslash}p{#1}}

\def\be{\begin{equation}}
\def\ee{\end{equation}}

\def\cc{{\rm cm}^{-3}}

\def\msun{{\rm M}_{\odot}}

\def\msunpc2{\msun/{\rm pc}^{2}}

\def\gsim{\lower.5ex\hbox{\gtsima}} 
\def\lsim{\lower.5ex\hbox{\ltsima}} 
\def\gtsima{$\; \buildrel > \over \sim \;$} 
\def\ltsima{$\; \buildrel < \over \sim \;$} \def\gsim{\lower.5ex\hbox{\gtsima}} 
\def\lsim{\lower.5ex\hbox{\ltsima}} 
\def\simgt{\lower.5ex\hbox{\gtsima}} 
\def\simlt{\lower.5ex\hbox{\ltsima}}

\def\nh2{n_{\rm H2}}

\definecolor{mkcolor}{HTML}{01abdf} 
\definecolor{apcolor}{HTML}{b3003b}
\definecolor{afcolor}{HTML}{01bdff}
\definecolor{blue-violet}{rgb}{0.54, 0.17, 0.89}
\definecolor{ao}{rgb}{0.0, 0.5, 0.0}
\definecolor{auburn}{rgb}{0.43, 0.21, 0.1}

\newcounter{subtab}[table]

\newcommand{\subtabcaption}[2][]{%
  \refstepcounter{subtab}%
  \captionof*{table}{\textbf{(\alph{subtab})} #2}%
  \if\relax\detokenize{#1}\relax\else\label{#1}\fi
}

\graphicspath{{plots/}}

\newcommand{\appref}[1]{\hyperref[#1]{Appendix~\ref*{#1}}}

\AtBeginDocument{%
  \newcommand{\Sectionref}[1]{\hyperref[#1]{Section~\ref*{#1}}}
}

\usepackage[switch]{lineno}
\nolinenumbers

\begin{document}

\title{Systematic selection of surrogate models for nonequilibrium chemistry}
\titlerunning{Systematic selection of surrogate models for nonequilibrium chemistry}


\author{
Robin Janssen\orcidlink{0009-0006-4356-3059} \inst{1}\inst{2} \and
Lorenzo Branca\orcidlink{0000-0002-6064-1964} \inst{1} \and 
Tobias Buck\orcidlink{0000-0003-2027-399X} \inst{1}
}

\authorrunning{Janssen et al.}

\institute{Interdisciplinary Center for Scientific Computing, Heidelberg University \and Institute of Computer Engineering, Heidelberg University}

\date{YYY; accepted XX XX, XXXX}

\abstract
{Nonequilibrium chemistry is central to many astrophysical environments, but remains a major computational bottleneck in simulations because solving the associated stiff, coupled, ordinary differential equation systems is expensive. Neural surrogates promise substantial increases in speed, yet most existing studies are limited to proof-of-concept demonstrations and lack rigorous, dataset-grounded comparisons of architectures or systematic optimization toward accuracy and efficiency.}
{We aim to establish a principled procedure for optimizing and selecting surrogate models for astrochemistry that would enable representative and quantitative comparisons across architectures. This requires joint optimization for accuracy and efficiency, suitable metrics for performance assessment, and evaluation of surrogate reliability under practical constraints such as uncertainty quantification (UQ) and iterative prediction.}
{To this end, we employed CODES, a benchmarking framework that performs multi-objective hyperparameter tuning, trains optimized configurations, and evaluates their behavior across multiple dimensions. We compared four surrogate families, two fully connected models and two latent-evolution models. These surrogates were optimized and trained on four \texttt{KROME}-generated datasets spanning primordial and molecular-cloud chemistry, with up to 287 reactions across 37 species, including parametric variations in radiation field and metallicity. Each model predicted chemical abundances and temperature over a 10~kyr interval for user-specified output times.}
{Dual-objective optimization reveals pronounced accuracy–efficiency trade-offs for all architectures and enables substantial efficiency gains with minimal accuracy loss. Across datasets, architectures group naturally by inductive bias: fully connected models, which impose minimal structural assumptions, achieve the highest accuracy and the most reliable UQ, but show the characteristic long-term error growth associated with low-bias models. Latent-evolution models -- though less accurate -- exhibit reduced error accumulation under iterative rollouts.}
{Our results underscore the importance of systematic optimization and comprehensive architectural comparison to make trade-offs explicit. The datasets, architectures, metrics, and benchmarking procedure are publicly bundled in CODES to support representative and reproducible comparisons.}


\maketitle
\section{Introduction}
\label{sec:intro}

Nonequilibrium chemistry plays a fundamental role in regulating a wide range of astrophysical processes, from the early Universe \citep{galli:1998,glover:2008}, to protoplanetary disks \citep{caselli:2012}, the evolution of galaxies \citep{pallottini:2017_b, lupi:2019}, and the star formation process \citep{decataldo:2019, kim:2018}.
In particular, accurately solving the chemical evolution in numerical simulations is crucial for modeling realistic thermodynamics. Directly solving the chemical evolution avoids the need for approximate treatments, such as precomputed analytical cooling functions, which may fail in dynamically evolving environments \citep{bovino19}. 

On the one hand, several codes implement extensive photo\mbox{}chemical networks and achieve high physical fidelity, but at significant computational cost. Examples include \texttt{CLOUDY} \citep{ferland:2017}, \texttt{UCLCHEM} \citep{holdship:2017}, and \texttt{MAIHEM} \citep{gray:2019}. This type of approach (see \citealt{olsen:2018} for a review) is typically used to post\mbox{-}process astrophysical simulations to obtain emission lines \citep[e.g.,][]{vallini:2018}. On\mbox{-}the\mbox{-}fly coupling of such detailed chemistry is therefore commonly limited to one-dimensional applications, such as photochemical evolution during shock processes \citep[e.g.,][]{danehkar:2022}.

On the other hand, some software packages are designed for coupling to full three-dimensional hydrodynamic simulations. \texttt{ASTROCHEM} \citep{kumar:2013MNRAS} and \texttt{KROME} \citep{grassi:2014} generate solver code for user-defined chemical networks, while \texttt{NIRVANA} \citep{ziegler:2016} and \texttt{GRACKLE} \citep{smith:2017} provide subroutines to solve selected, astrophysically motivated networks. More recently, \texttt{CARBOX} \citep{carbox}, the first fully differentiable astrochemical code written entirely in JAX \citep{jax2018github}, was introduced. It resembles previous approaches but exposes the full computational graph (reaction rates, Jacobians, and state updates), enabling direct access to model internals and making fast, gradient-based inference on external parameters feasible.

However, solving the full system of ordinary differential equations (ODEs) describing chemical networks on the fly remains one of the dominant computational bottlenecks in hydrodynamical simulations.
This occurs for several reasons: (i) astrochemical networks contain reactions with rate coefficients that span many orders of magnitude. Consequently, the associated chemical relaxation times can be much shorter than, comparable to, or longer than the local hydrodynamical or advection time depending on density, temperature, and irradiation. Thus, while some channels drive very fast transients (often far faster than the concurrent fluid evolution), equilibrium cannot be assumed globally or at all times; (ii) the governing ODEs are stiff and strongly coupled \citep{branca:2023, nejad:2005}, which necessitates costly implicit integrators; (iii) the number of reactions grows super-linearly with the number of chemical species, inflating the required floating\mbox{-}point operations; and (iv) in massively parallel simulations, load imbalance and communication or synchronization overheads reduce effective throughput.

Recent advances in deep learning have enabled the development of surrogate models that aim to mitigate these bottlenecks by replacing numerical solvers with computationally cheaper, typically data-driven architectures. A variety of surrogate approaches have been proposed, including latent ODEs with autoencoders \citep{grassi:2021, holdship:2021}, physics-informed neural networks \citep{branca:2023}, neural ODEs with autoencoders \citep{sulzer:2023}, neural fields \citep{asensio24}, parametric neural ODEs \citep{gijs25}, and neural operators \citep{branca24, pelle25, ono25}.

Compared to numerical solvers, neural surrogates are often more efficient because they transform the computational workload into forms that map  to modern accelerator hardware well. Inference in neural networks is dominated by dense linear algebra operations, which are highly parallel and can be executed efficiently on GPUs, often at reduced numerical precision. Depending on the architecture, surrogate inference additionally has a fixed computational cost per evaluation, in contrast to the adaptive behavior of ODE solvers. This predictability simplifies load balancing and improves throughput in large-scale simulations.
However, these efficiency gains come at the cost of approximation error. It may be for this reason that up to now surrogate models have not been widely adopted in current state-of-the-art astrophysical simulations. Developing surrogate models for this application context is challenging; chemical solvers are invoked at every hydrodynamic time step, which means that even small inaccuracies in surrogate predictions can lead to significant errors through numerical instability and error accumulation. 

To avoid this error build-up, surrogate models generally require a high level of accuracy, but there are additional factors to be taken into account when determining the optimal surrogate architecture and configuration for a task; a good average accuracy may not be sufficient if the surrogate occasionally produces catastrophic errors. Errors may be more consequential for some predictions than for others, the prediction interval must often be flexible due to adaptive hydrodynamic time stepping, and a reliable uncertainty quantification (UQ) mechanism is desirable to enable fallback to the numerical solver when predictions are untrustworthy. Given these (and possibly additional) desirables, we argue that advancing from proof-of-concept models to simulation-ready tools requires a paradigm shift in how surrogate models for nonequilibrium chemistry are developed and evaluated. To facilitate this shift, we developed the CODES benchmark \citep{codes_2024}. 

The remainder of this paper is structured as follows. In Sect. \ref{sec:data} we describe the astrochemical datasets used for training and evaluation. Section \ref{sec:surrogates} introduces the surrogate architectures and their variants, while Sect. \ref{sec:CODES} details the benchmarking, optimization, and evaluation framework. The results are presented in Sect. \ref{sec:results}, followed by a discussion in Sect. \ref{sec:discussion} and conclusions in Sect. \ref{sec:conclusion}.

\section{Chemistry datasets}
\label{sec:data}

\begin{table}
\caption{Parameter ranges used to generate the datasets in this work.}
\centering
\begin{tabular}{l|lll }
 \toprule
 Quantity         & Variable          & Min & Max \\ 
 \midrule
 Gas density      & $\log (n/\cc)$     & -2 & 3.5 \\  
 Abundance ($\rm H$ and $\rm He$)        & $\log (n_i/n)$     & -6 & 0   \\  

 Abundance (metals)  & $\log (n_i/n)$ & -20 & -6 \\
 Temperature      & $\log (T/\rm K)$    & $\log(20)$  & 5.5\\    
 Radiation        & $\log(G/G_0)$     & -15 & -5 \\

 Metallicity      & $\log(Z/Z_{\odot})$ & -3 & 0 \\
 Time             & $\log(t/\rm yr)$        & -1 & 4 \\
 \bottomrule
\end{tabular}
\tablefoot{All datasets vary the initial gas density ($n$), species abundances ($n_i/n$), and temperature ($T$), while parametric datasets additionally vary the radiation intensity ($G)$ and metallicity $(Z)$ (see Sect. \ref{sec:data}). 
The datasets contain 2048 (primordial), 4096 (primordial parametric and cloud), and 8192 (cloud parametric) initial conditions, each evaluated at 100 time steps.}
\label{tab:data_set_structure}
\end{table}
The synthetic data used for training, validation and testing of the surrogate models was generated with \texttt{KROME} \citep{grassi:2014}, a \texttt{Python} preprocessor that produces \texttt{Fortran} code to solve a given chemical network. The system of ODEs that describes the time evolution of the chemical species is

\begin{equation}
\dot{n}_k = \sum_{i,j} A^{ij}_k \, n_i \, n_j + \sum_{i} B_k^i \, n_i\,,
\label{2body}
\end{equation}

\noindent where the coefficients $A^{ij}_k = A^{ij}_k(T,\mathbf{n})$ and $B_k^i = B_k^i(\mathbf{F})$\footnote{Although the coefficients $B^i_k$ can generally depend on gas temperature, they do not for the reaction networks considered here.} denote net reaction contributions to species $k$.
In particular, while the underlying rate coefficients are non-negative, $A^{ij}_k$ and $B_k^i$ include the appropriate stoichiometric numbers (with their signs) so that they account for both the production and destruction of $n_k$.
Here $T$ is the gas temperature, $\mathbf{n}$ is the vector of species abundances, and $\mathbf{F}$ denotes any ionizing or dissociation flux.
The system is closed by simultaneously evolving the gas temperature ($T$):
\begin{equation}
    \label{temp_evolution}
    \dot{T}=\frac{(\gamma-1)}{k_b\sum_{i}n_i}(\Gamma-\Lambda)\,,
\end{equation}
where $k_b$ is the Boltzmann constant, $\gamma$ is the gas adiabatic index, $\Gamma=\Gamma(T,\mathbf{n},\mathbf{F})$ and $\Lambda=\Lambda(T,\mathbf{n},\mathbf{F})$ are the heating and cooling functions, respectively\footnote{Equations \ref{2body} and \ref{temp_evolution} may, in principle, also depend on additional external parameters, such as the metallicity $(Z)$ or the rate of molecule formation on dust grains; however, for clarity of notation, these dependencies have been omitted.}. To solve Eqs.~\ref{2body} and \ref{temp_evolution}, \texttt{KROME} uses a backward differentiation formula solver of \texttt{DLSODES} \citep{odepack}, an implicit multistep solver that exploits the sparsity of the Jacobian matrix constructed with the chemical fluxes.\footnote{We adopt the default errors for \texttt{KROME}, i.e., relative and absolute tolerances are fixed at $10^{-6}$ and $10^{-20}$ respectively.}

To demonstrate the performance of our models, we constructed four datasets with distinct characteristics: (\textit{i}) primordial, (\textit{ii}) primordial parametric, (\textit{iii}) cloud, and (\textit{iv}) cloud parametric.
In this context, parametric refers to datasets where the sampling varies not only over the initial conditions but also over certain external parameters. This distinction is important because these parameters constitute additional, structurally distinct inputs that may require architectural modifications (see Sect. \ref{sec:surrogates}). Specifically, we varied two physical parameters: the radiation field intensity ($G$) and the metallicity ($Z$). For the primordial and primordial parametric datasets, we considered 9 chemical species: $\mathrm{e}^-$, $\mathrm{H}^-$, $\rm H$, $\mathrm{H}^+$, He, $\mathrm{He}^+$, $\mathrm{He}^{++}$, $\mathrm{H}_2$, and $\mathrm{H}_2^+$. These species evolve through 46 reactions\footnote{The reaction rates are taken from \citet{bovino:2016}: reactions 1 to 31, 53, 54, and from 58 to 61 in their Tables B.1 and B.2, photo-reactions P1 to P9 in their Table 2.}, including H$_2$ formation on dust grains \citep{jura:1975, wakelam:2017}, photo-chemistry, and cosmic-ray ionization. For the cloud and cloud parametric datasets, we extended the primordial network with heavier atoms and molecules:
$\mathrm{C^-}$, $\mathrm{O^-}$, $\mathrm{C}$, $\mathrm{Si}$, $\mathrm{O}$,
$\mathrm{OH}$, $\mathrm{CO}$, $\mathrm{CH}$, $\mathrm{CH_2}$, $\mathrm{C_2}$, $\mathrm{HCO}$, $\mathrm{H_2O}$, $\mathrm{O_2}$,
$\mathrm{C^+}$, $\mathrm{Si^+}$, $\mathrm{O^+}$, $\mathrm{HOC^+}$, $\mathrm{HCO^+}$, $\mathrm{H_3^+}$, $\mathrm{CH^+}$, $\mathrm{CH_2^+}$, $\mathrm{CO^+}$,
$\mathrm{CH_3^+}$, $\mathrm{OH^+}$, $\mathrm{H_2O^+}$, $\mathrm{H_3O^+}$, $\mathrm{O_2^+}$, and $\mathrm{Si^{2+}}$, resulting in a network of 37 species and 287 reactions\footnote{The reactions rate are taken from \citet{Glover2010}, \citet{Glover2011}, \citet{KF1996}, \citet{OConnor2015}, \citet{Glover2009} and KIDA \citep{wakelam:2012}, OSU \citep{OSU1980}, and UMIST \citep{UMIST1991} databases.}.

In the case of the nonparametric datasets, the ionizing radiation intensity is fixed at $G = G_0$, where $G_0 = 1.6 \times 10^{-3} \ \mathrm{erg \ cm^{-2} \ s^{-1}}$, and the metallicity is set to the solar value, $Z = Z_{\odot}$.
For the parametric datasets, the radiation intensity varies over the range $G \in [0.1 \ G_0,\ 10 \ G_0]$, and the metallicity spans the interval $Z \in [10^{-3} \ Z_{\odot},\ Z_{\odot}]$\footnote{Because the chemical networks differ between the datasets, the interpretation of “metallicity” and the corresponding cooling treatment also differs. In the primordial case, metallicity denotes the total abundance of all elements heavier than helium and is implemented following the prescription of \citet{shen:2013}. In the cloud case, by contrast, metallicity is applied by rescaling the relative abundances of heavy atoms and molecules with respect to hydrogen and helium.}. 
In all datasets, we assume a cosmic-ray ionization rate $\zeta_{\mathrm{cr}} = 3 \times 10^{-17}\, \mathrm{s}^{-1}$ and a dust to gas ratio in solar units $f_d=0.3$.

A further critical aspect in constructing representative datasets for training surrogate models is the choice of how to sample the initial parameter space. For surrogate training, the primary objective is to obtain a space-filling design that broadly covers the admissible domain for abundances and potential additional parameters, rather than to reproduce an assumed astrophysical prior distribution. The dimensionality of this domain ranges from ten parameters in the primordial case to 39 in the cloud parametric case. We therefore employed Sobol sampling \citep{Sobol1967}, a deterministic quasi-random sampling strategy that provides near-uniform coverage of bounded parameter domains. Compared to standard random sampling, Sobol sequences reduce clustering and uncovered regions at fixed sample size, improving representation of rare parameter combinations while remaining reproducible. Sobol sampling is a global, non-adaptive strategy and therefore does not focus additional samples on regions where the surrogate model exhibits larger prediction errors--addressing such regions would require an active learning or adaptive sampling strategy.

After drawing a point in parameter space, we construct the initial abundances $\mathbf{n}_0$ by enforcing a helium mass fraction consistent with a 3:7 He-to-(H+He) ratio, and we compute the electron abundance a posteriori to ensure charge neutrality. In the cloud parametric case, all metal-bearing species are initialized by scaling a reference abundance pattern linearly with the sampled metallicity $Z$ (i.e. $n_X \propto Z$ for heavy elements), while H/He species follow the helium constraint above. The thermochemical evolution is then computed for $10~\mathrm{kyr}$ using \texttt{KROME}. The resulting trajectories are sampled at 100 logarithmically spaced time points starting just below $0.1~\mathrm{yr}$ to resolve the stiff early-time dynamics. The ranges of initial conditions and parameters are summarized in Table \ref{tab:data_set_structure}.

\section{Surrogate architectures}
\label{sec:surrogates}

\begin{figure*}
  \centering
  \includegraphics[width=\textwidth]{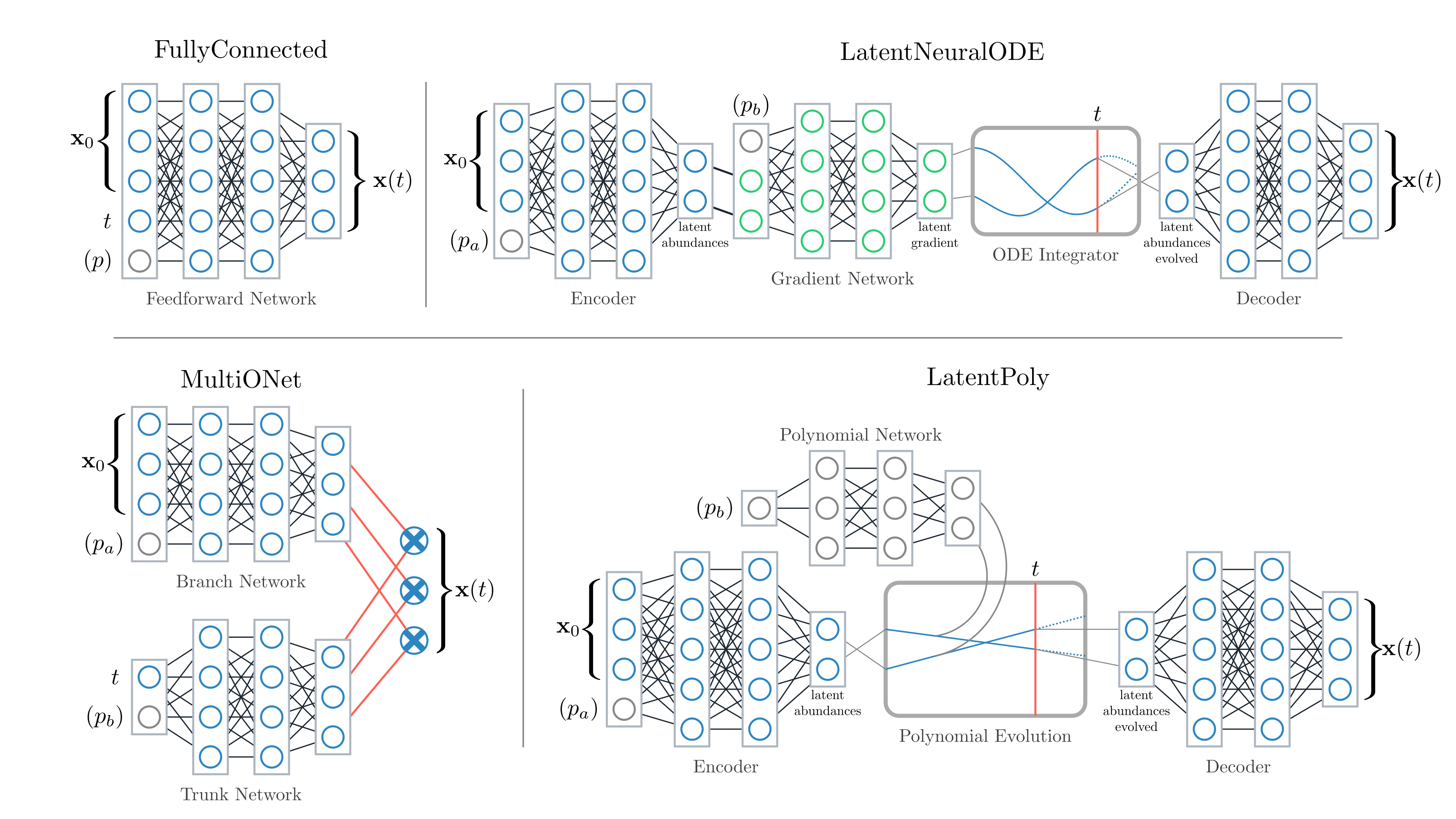}
  \caption{Schematic overview of the four surrogate architectures investigated in this work. \textit{Left:} Fully connected surrogates (\texttt{FCNN}, \texttt{MON}). \textit{Right:} Latent-evolution surrogates (\texttt{LNODE}, \texttt{LP}). All models take the initial state ($\mathbf{x}_0$), the desired output time ($t$), and, for parametric datasets, additional physical parameters ($p$), and output the predicted state $\mathbf{(x}(t))$. Optional components are shown in gray, and parameter-handling options ($p_a$ and $p_b$) denote alternative ways of incorporating the parameters.}
  \label{fig:archs}
\end{figure*}

The surrogate architectures investigated in this work are implemented within our benchmarking framework CODES, which is described further in Sect. \ref{sec:CODES}. To ensure fair comparison, reproducibility, and extensibility, all architectures follow a mandatory code structure. For example, all surrogates implement a common data preprocessing and training interface and expose a unified prediction method. Design choices were further guided by the target application: replacing the numerical solver used to evolve chemical abundances across a hydrodynamical time step. Since the hydrodynamical time step can vary during a simulation, all surrogates are trained to predict the system state at user-specified times within the temporal range covered by the training data, given an initial condition. During training, trajectories are sampled at fixed, logarithmically spaced time points, so that the time variable takes only discrete values in the training set; however, time is treated as a continuous input by the architectures, allowing predictions at intermediate times within the trained interval.

CODES currently includes four surrogate architectures:

\begin{itemize}
\item \texttt{FullyConnected} (\texttt{FCNN}). A standard fully connected neural network. Initial abundances and the desired output time are concatenated and provided jointly as input to the model.

\item \texttt{MultiONet} (\texttt{MON}). An adaptation of DeepONet \citep{lu_learning_2021} for multiple outputs as proposed in \cite{lu_2022_comprehensive}. The inputs to the branch network are the initial abundances, while the trunk network receives the desired output time.

\item \texttt{LatentNeuralODE} (\texttt{LNODE}). Following \cite{sulzer:2023}, an autoencoder is combined with a latent space neural ODE \citep{chen_neural_2018}. The encoder receives the initial abundances as input. In contrast to \texttt{FCNN} and \texttt{MON}, this model features an explicit time evolution mechanism, hence the time input is used as the endpoint of the numerical integration, not as an explicit network input.

\item \texttt{LatentPoly} (\texttt{LP}). An autoencoder with a learnable latent space polynomial, also following \cite{sulzer:2023}. Similarly to \texttt{LNODE}, the initial state is input to the encoder network, and the time input controls the evaluation point of the learnable polynomial.
\end{itemize}

Schematics of all architectures are displayed in Fig. \ref{fig:archs}. Generally, the surrogate architectures investigated in this work are not autoregressive: the desired output time $t$ is provided explicitly, and the models learn to predict the system state at arbitrary times for a given initial condition. This design enables flexible prediction intervals required for simulations with adaptive hydrodynamical time stepping. All architectures accept a continuous time input $t$, allowing predictions at arbitrary times within the training horizon (10~kyr in this work).

As mentioned in Sect. \ref{sec:data}, the ODE governing the evolution of a chemical species may depend not only on the abundances of other species, but also on additional physical quantities. One important such quantity is the temperature ($T$), which coevolves with chemical abundances. Due to this explicit time-dependence, $T$ is treated as an additional species in the chemical reaction network. Other physical quantities, such as the radiation field intensity ($G$) and the metallicity ($Z$), are updated independently of the chemistry and therefore remain constant over a hydrodynamical time step.
In principle, these quantities could also be treated as additional species that stay constant across one set of trajectories. In this view, we simply concatenate the parameters to the abundances and hand the combined vector to the architecture as input. However, since these parameters are structurally and physically distinct from chemical abundances, we propose architectural variations that better reflect this difference for all architectures except \texttt{FCNN}\footnote{The motivation to include \texttt{FCNN} as a surrogate architecture is to simply hand all available information to the model without any additional inductive bias. Due to this, it did not seem apt to derive an architectural variation for the parameters for this model. The variations for the other architectures are structurally well motivated, but for \texttt{FCNN}, we feel that this would not be the case.}. 

For \texttt{MON}, the parameters can be handed to either the branch or the trunk network. The latter approach is in line with the original motivation to have two separate networks, namely the fact that the inputs to the branch and trunk network have different domains. For \texttt{LNODE}, the alternative to handing parameters to the encoder network (concatenated with the abundances) is to hand them to the gradient network (concatenated to the encoded latent-space abundances). The intuitive reading of this would be that while the abundances should be encoded to evolve on a lower-dimensional latent manifold, the parameters might be more aptly viewed as influencing the gradients in this latent space, rather than changing the projection into this space. Finally, for \texttt{LP}, we propose as alternative to the joint encoding of abundances and parameters that the parameters should influence the learnable polynomial coefficients directly. This is achieved by introducing an additional fully connected neural network (the polynomial network), which predicts the polynomial coefficients based on the parameter values. These architectural variations are treated as optional hyperparameters and are only used when selected during hyperparameter tuning.

\section{Methodology: Benchmarking and optimization framework}  
\label{sec:CODES}

Initially developed to compare surrogate models on a given dataset, the CODES (\textbf{C}oupled \textbf{ODE} \textbf{S}urrogates) benchmark \citep{codes_2024} has evolved into a general framework for optimizing and evaluating surrogate architectures for coupled ODE systems\footnote{In terms of data structure, the only assumption made in CODES is that trajectories evolve with respect to an independent variable, but some architectures' inductive bias is aimed specifically at ODE data.}. Optimization is achieved through extensive hyperparameter tuning, while evaluation consists of training multiple instances of each optimized architecture on varying subsets of the dataset and assessing their performance across multiple dimensions. The result of this evaluation step is a collection of metrics and diagnostic plots aimed at both quantifying surrogate performance and comparing architectures. These metrics and plots extend beyond standard measures of accuracy to address the above-mentioned desirables, yielding detailed insights into the computational demand, error distribution, generalisation capabilities, and the reliability of UQ.

CODES is implemented in \textsc{PyTorch} \citep{paszke_pytorch_2019} and builds on established tools including \textsc{Optuna} \citep{akiba2019optuna} for hyperparameter optimization (HPO), \textsc{schedulefree} \citep{defazio_road_2024} for learning-rate scheduling, and the \textsc{torchode} \citep{lienen2022torchode} package for differentiable GPU-based ODE solvers. The code is available on GitHub\footnote{\href{https://github.com/AstroAI-Lab/CODES-Benchmark}{https://github.com/AstroAI-Lab/CODES-Benchmark}.}. CODES follows standard software best practices, including automated testing, documentation, and reproducibility via controlled seeding and stored configuration files. In the following, we introduce the three-step process of hyperparameter tuning, training and evaluation, and elaborate on metrics and UQ.

\begin{figure*}
  \centering
  \includegraphics[width=.9\textwidth]{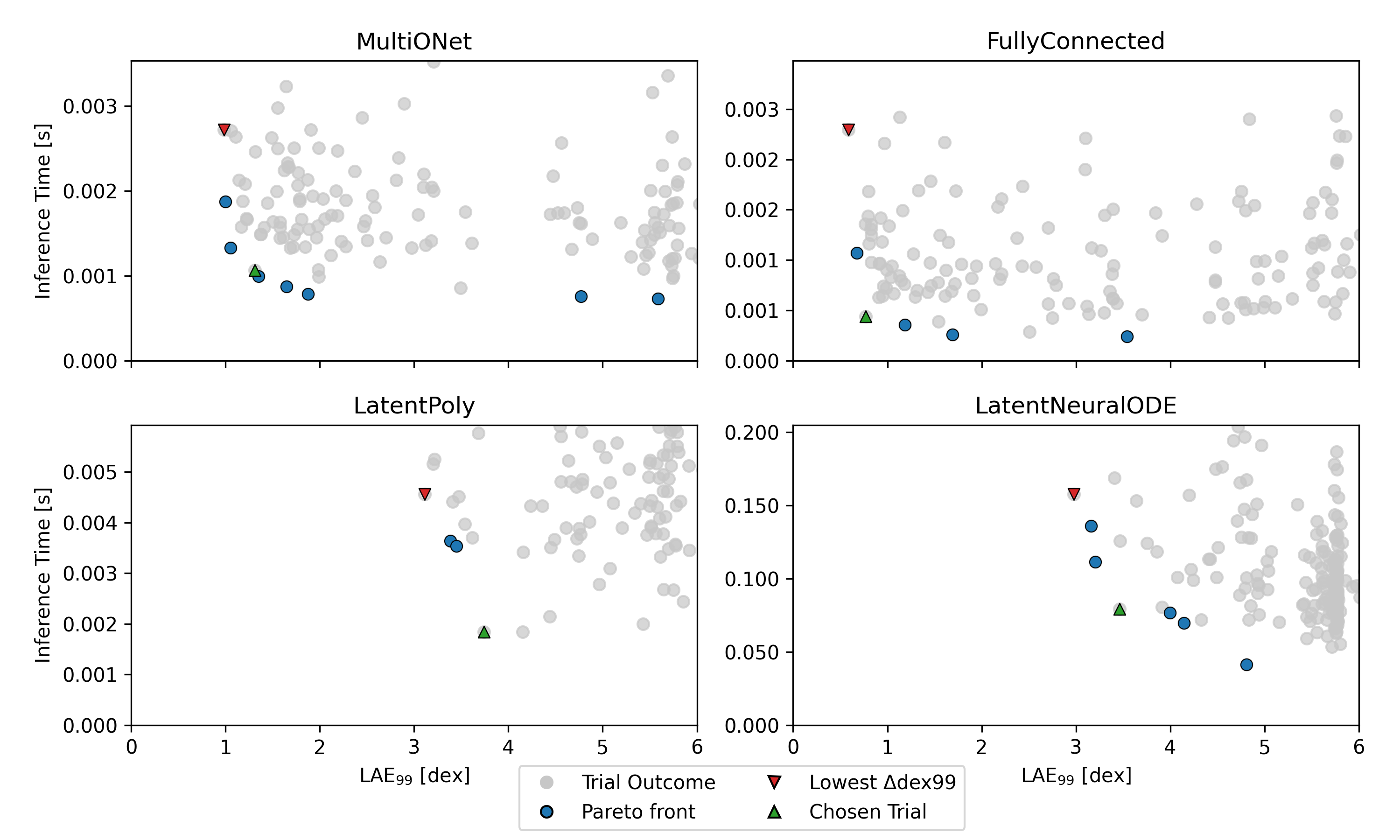}
\caption{Pareto fronts obtained from dual-objective hyperparameter tuning for the primordial dataset. 
The colored points indicate Pareto-optimal configurations, while the gray points are dominated solutions. 
The red marker denotes the lowest-error configuration, and the green marker indicates the selected trade-off configuration balancing accuracy and inference time.}
  \label{fig:pareto_fronts}
\end{figure*}

\subsection{Multi-objective hyperparameter tuning}
\label{subsec:tuning}

Hyperparameter optimization is a necessary first step in the benchmarking pipeline, as meaningful comparisons require well-performing configurations and the performance of neural networks strongly depends on the chosen hyperparameters. Compared to its initial formulation in \cite{codes_2024}, CODES was extended to replace the assumption of known optimal hyperparameters with an explicit HPO pipeline. To clarify terminology, a trial denotes a single training run within HPO. In each trial, the model is trained on the training set using the selected hyperparameters, and its performance is subsequently evaluated on the validation set.  A study corresponds to the optimization of one surrogate architecture on one dataset and thus comprises multiple trials. A tuning refers to the joint optimization of all architectures on a given dataset. In this work, each tuning consisted of four studies, hence across the four datasets considered, a total of sixteen studies were performed. Given the high computational cost of the hyperparameter tuning runs, we rely on \textsc{Optuna} to efficiently explore the hyperparameter space. CODES leverages \textsc{Optuna}’s adaptive sampling strategies, early stopping of unpromising trials, and parallel execution to reduce computational overhead while maintaining robust optimization performance. The specific samplers and pruning strategies used in this work are described in Appendix \ref{app:hyperparams:setup}.

An optimization procedure requires the definition of an objective function. For the results presented in Sect. \ref{sec:results}, we performed dual-objective optimization, simultaneously targeting prediction accuracy and computational efficiency. Accuracy is quantified using the LAE$_{99}$, while efficiency is measured via inference time, defined as the wall-clock duration required for a complete forward pass over all samples\footnote{Inference time is measured on the validation set during HPO and on the test set in the final evaluation.}. The choice and interpretation of these metrics are discussed in detail in Sect. \ref{subsec:metrics}.
Each trial is thus associated with two objective values, LAE$_{99}$ and inference time. To handle this dual-objective setting, CODES employs \textsc{Optuna}’s NSGA-II evolutionary algorithm \citep{deb_fast_2002}, which is designed to identify a set of Pareto-optimal solutions. 
A configuration is Pareto-optimal if no other trial achieves strictly better performance in both accuracy and efficiency simultaneously, and the collection of such configurations forms a Pareto front in the two-dimensional objective space. NSGA-II is well suited for this setting, as it directly supports multi-objective optimization in heterogeneous search spaces and promotes broad exploration.

Examples of the resulting Pareto fronts for the primordial dataset are shown in Fig. \ref{fig:pareto_fronts}. From each Pareto front, we manually selected a configuration near the knee of the curve, balancing accuracy and efficiency\footnote{In principle, this step could be automated, but in practice such automation offers little benefit. Manual selection is more robust given the stochastic nature of the tuning. Further details are given in Appendix \ref{app:hyperparams}.}. The number of trials per study is set heuristically to $15 \times N_h$, where $N_h$ (ranging from 11 to 17) denotes the number of tuned hyperparameters. The hyperparameters optimized this way fall into two categories: architecture-specific and shared hyperparameters. The architecture-specific parameters differ between models and include latent dimensionalities, hidden-layer widths, or choices governing how external parameters are incorporated. The shared hyperparameters govern aspects of the training procedure and are common across architectures. The resulting search space is heterogeneous and includes real-valued, categorical, and conditionally activated parameters. We refer to Appendix \ref{app:hyperparams} for further details on the tuning procedure and a full list of optimized hyperparameters.

\subsection{Training and evaluation protocol}
\label{subsec:train_eval}

Following HPO, all surrogate architectures were trained under standardized conditions using the optimized hyperparameters to enable fair comparison across datasets. For most evaluations, it is sufficient to train one instance of the surrogate architecture on the full training dataset (the main model). For additional characterization, CODES provides five modalities, each of which can be toggled via a central configuration file and results in training extra models on modified datasets:

\begin{itemize}
    \item Interpolation. Tests a surrogate’s ability to interpolate between training time steps by omitting intermediate points.
    
    \item Extrapolation. Probes how well a model generalizes beyond the training time range by truncating trajectories.  
    
    \item Sparsity. Assesses robustness when the number of training samples is reduced.
    
    \item Batch size. Examines the effect of training with different batch sizes.  
    
    \item Uncertainty quantification. Estimates predictive uncertainty with deep ensembles (DEs; \citealt{lakshminarayanan_simple_2017}).
\end{itemize}

All modalities except UQ support setting multiple parameter settings\footnote{Deep ensembles require multiple models by definition, hence there is only a single setting for UQ: the number of models in the DE.}, enabling the construction of scaling curves that show how performance changes as interpolation gaps widen, trajectories are truncated earlier, training data grows sparse, or batch size decreases. These modalities serve as diagnostic tools for probing surrogate robustness under controlled perturbations of the training setup. In this work, we focused on UQ, as it is most directly relevant to assessing surrogate reliability in simulation settings.

Upon completion of the training step, architectures are evaluated on the test set. 
The evaluation procedure reloads the trained weights and computes a core set of quantitative performance metrics for the main model and all additional models trained under the enabled modalities. Depending on the configuration, further diagnostics can be enabled, including loss trajectories, correlations between prediction errors and true gradients, inference times, and the computational footprint of each architecture (trainable parameters and memory usage). The results of these evaluations form the basis for the comparative plots and tables shown in Sect. \ref{sec:results}.

\subsection{Accuracy metrics}
\label{subsec:metrics}

Many standard accuracy metrics are unsuitable to assess model performance in the context of astrochemistry, where chemical abundances spread over many orders of magnitude (in our datasets up to 30 dex). On such datasets, absolute error metrics such as the mean absolute error or the mean squared error do not account for errors on scarce quantities. In principle, relative error metrics such as the mean relative error mitigate this problem by dividing the absolute error by the actual magnitude of the quantity in question. However, when aggregating relative error distributions, a small number of disproportionately large errors can strongly skew summary statistics, especially the mean. Furthermore, relative errors are not symmetric -- significantly overestimating an abundance produces large relative errors, whereas significantly underestimating it produces at most a relative error of 100\% -- and division by very small values can be numerically unstable. 

These problems with relative errors could be mitigated by introducing a relative error threshold value, which would serve as a stabilizing factor in the division, preventing explosive error values and numerical instabilities. This, however, implicitly introduces a weighting of abundances, as errors below the chosen threshold are effectively discounted. While weighting errors is a viable approach, this weighting should be motivated by the actual importance of the abundance to the evolution of the system, which is hard to quantify and presumably not captured adequately by a blanket threshold value. In the present work, all species and time steps are treated equally when aggregating error statistics, although CODES may provide custom weighting schemes in future extensions.

For these reasons, we propose to focus on log-space metrics when measuring accuracy of surrogate models in astrochemistry (i.e., across HPO, training and evaluation). All models in this work are trained on log-transformed abundances, hence it is natural to compute aggregate values, losses and optimization objectives in log-space as well. Concretely, we report the models' log-space mean absolute error (mLAE) as well as the LAE$_{99}$ in Table \ref{table:results}, both computed on the $\text{log}_{10}$-transformed abundances. As mentioned in Sect. \ref{subsec:tuning}, the latter metric was also used as optimization objective during HPO\footnote{Due to the usage of $\text{log}_{10}$, the metrics indicate errors in orders of magnitude and hence have the unit "dex."}. We deliberately employed a percentile-based tail metric rather than the maximum error. Optimizing directly for low maximum errors directs computational effort toward eliminating rare failure cases rather than achieving good performance in the majority of cases. Percentile-based metrics accept rare failures while still capturing overall predictive performance, which is consistent with the use of a UQ-enabled solver fallback. The choice of the 99th percentile is pragmatic -- ideally, this threshold should be motivated by the requirements of the target application.

\begin{table*}[t]
\centering
\caption{\centering Performance metrics across datasets and models. }
\label{table:results}
\small
\setlength{\tabcolsep}{0pt}
\renewcommand{\arraystretch}{1.0}
\begin{tabular}{p{1.6cm} p{1.4cm} 
                C{1.4cm} C{1.4cm} C{1.4cm} C{2.0cm} C{1.4cm}
                C{1.6cm} C{1.6cm} C{1.6cm} C{1.3cm}}
\toprule
Dataset & Arch & mLAE & LAE$_{99}$ & mRE & Inference time & Params & mLAE$_\text{iter}$ & mLU$_\text{DE}$ & mLAE$_\text{DE}$ & UQ \\
 & & [dex] & [dex] & [\%] & [µs] & [k] & [dex] & [dex] & [dex] & PCC \\
\midrule
Primordial & MON   & 0.343 & 2.47 & 38.5 & 703 ± 77 & 498 & 0.499 & 0.225 & 0.174 & 0.531  \\
           & FCNN  & \textbf{0.094} & \textbf{0.693} & \textbf{11.8} & \textbf{307 ± 38} & 896 & \textbf{0.273} & 0.154 & \textbf{0.101} & \textbf{0.581}  \\
           & LNODE & 0.661 & 3.04 & 80.6 & 18190 ± 960 & 1418 & 1.26 & 0.251 & 0.686 & 0.267 \\
           & LP    & 0.843 & 3.35 & 90.8 & 619 ± 78 & \textbf{461} & 1.24 & 0.301 & 0.833 & 0.263\\
\midrule
Primordial & MON   & 0.613 & 5.96 & 41.6 & 2030 ± 90 & 86 & \textbf{0.667} & 0.431 & 0.251 & \textbf{0.695} \\
Parametric & FCNN  & \textbf{0.259} & \textbf{1.74} & \textbf{33.4} & \textbf{642 ± 42} & 454 & 0.719 & 0.172 & \textbf{0.121} & 0.512 \\
           & LNODE & 0.693 & 3.32 & 81.7 & 41520 ± 2360 & 409 & 1.58 & 0.821 & 0.783 & 0.278 \\
           & LP    & 0.743 & 3.43 & 85.4 & 1040 ± 120 & \textbf{27} & 1.60 & 0.488 & 0.805 & 0.425 \\
\midrule
Cloud & MON   & \textbf{0.034} & \textbf{0.326} & 4.1 & 13010 ± 10910 & 459 & \textbf{0.187} & 0.031 & \textbf{0.024} & 0.654 \\
      & FCNN  & 0.035 & 0.384 & \textbf{3.5} & \textbf{814 ± 98} & \textbf{24} & 0.246 & 0.111 & 0.054 & \textbf{0.796} \\
      & LNODE & 1.92 & 4.55 & 100.0 & 43430 ± 610 & 286 & 1.92 & 0.008 & 1.91 & 0.122 \\
      & LP    & 1.64 & 4.70 & 99.9 & 1910 ± 190 & 193 & 1.78 & 0.306 & 1.62 & 0.023 \\
\midrule
Cloud        & MON   & \textbf{0.121} & \textbf{0.831} & 16.3 & 8470 ± 4090 & 372 & 0.938 & 0.077 & \textbf{0.066} & 0.559 \\
Parametric   & FCNN  & 0.244 & 2.80 & \textbf{15.0} & \textbf{1510 ± 130} & 361 & \textbf{0.441} & 0.144 & 0.083 & \textbf{0.82} \\
             & LNODE & 1.51 & 4.80 & 99.7 & 49070 ± 520 & 234 & 1.85 & 0.197 & 1.49 & -0.017 \\
             & LP    & 1.37 & 4.51 & 99.7 & 1820 ± 120 & \textbf{17} & 2.04 & 0.679 & 1.40 & 0.051 \\
\bottomrule
\end{tabular}
\end{table*}

\subsection{Uncertainty quantification}
\label{subsec:uq_theory}

Even with a highly representative dataset, a surrogate model will still make significant errors for some fraction of its predictions. A well-calibrated UQ mechanism serves to detect these unreliable predictions and can enable a fallback to the numerical solver for such cases. To investigate UQ and its potential for error mitigation, CODES optionally trains a DE \citep[][]{lakshminarayanan_simple_2017} with a configurable number of models. DE rests on the idea that introducing some stochasticity\footnote{In CODES, this stochasticity is implemented by assigning each model a different seed for the random number generators, which influences the weight initialization and dataset shuffling during training.} into the training of multiple, architecturally identical models will cause them to converge to different, yet equally representative points on the loss landscape. The ensemble's prediction is obtained as the mean across ensemble member predictions, while the standard deviation of these predictions can be used as a measure of uncertainty. In the following, we refer to this standard deviation as the predicted uncertainty. 

For the evaluations in this work, the ensembles always consist of $M = 5$ models. Employing a DE comes at additional computational cost, because predictions require $M$ forward passes rather than one. Most surrogates investigated in this work are computationally lightweight, hence this overhead remains minor. Nevertheless, the additional expenditure must be justified by performance gains. The DE could deliver two main benefits: improving accuracy over the single-model setting, and providing a reliable identification mechanism for significant errors. In Sect. \ref{subsec:UQ}, we investigate whether these benefits are achieved in practice.

\section{Results}
\label{sec:results}

In the following, we present the results of applying CODES to the four datasets primordial, primordial parametric, cloud, and cloud parametric, described in  Sect. \ref{sec:data}. We begin with an analysis of the hyperparameter tuning runs in Sect. \ref{subsec:hyperparam_results}. The benchmark results following HPO are then examined from several perspectives: prediction accuracy (Sect. \ref{subsec:accuracy}), UQ in application-relevant settings (Sect. \ref{subsec:UQ}), and error propagation under iterative prompting (Sect. \ref{subsec:error_prop}). Key metrics are shown in Table \ref{table:results}, although this is only an excerpt of the full list of metrics generated by CODES.

\subsection{Hyperparameter tuning}
\label{subsec:hyperparam_results}

Following the order of the benchmarking pipeline, we begin by evaluating the dual-objective optimization procedure. The goal of this evaluation is not to analyze individual hyperparameters, but to investigate whether the hyperparameter tuning procedure proposed in Sect. \ref{subsec:tuning} yields meaningful accuracy–efficiency trade-offs and how these compare to accuracy-only optimization.

\begin{figure}
  \centering
\includegraphics[width=0.5\textwidth]{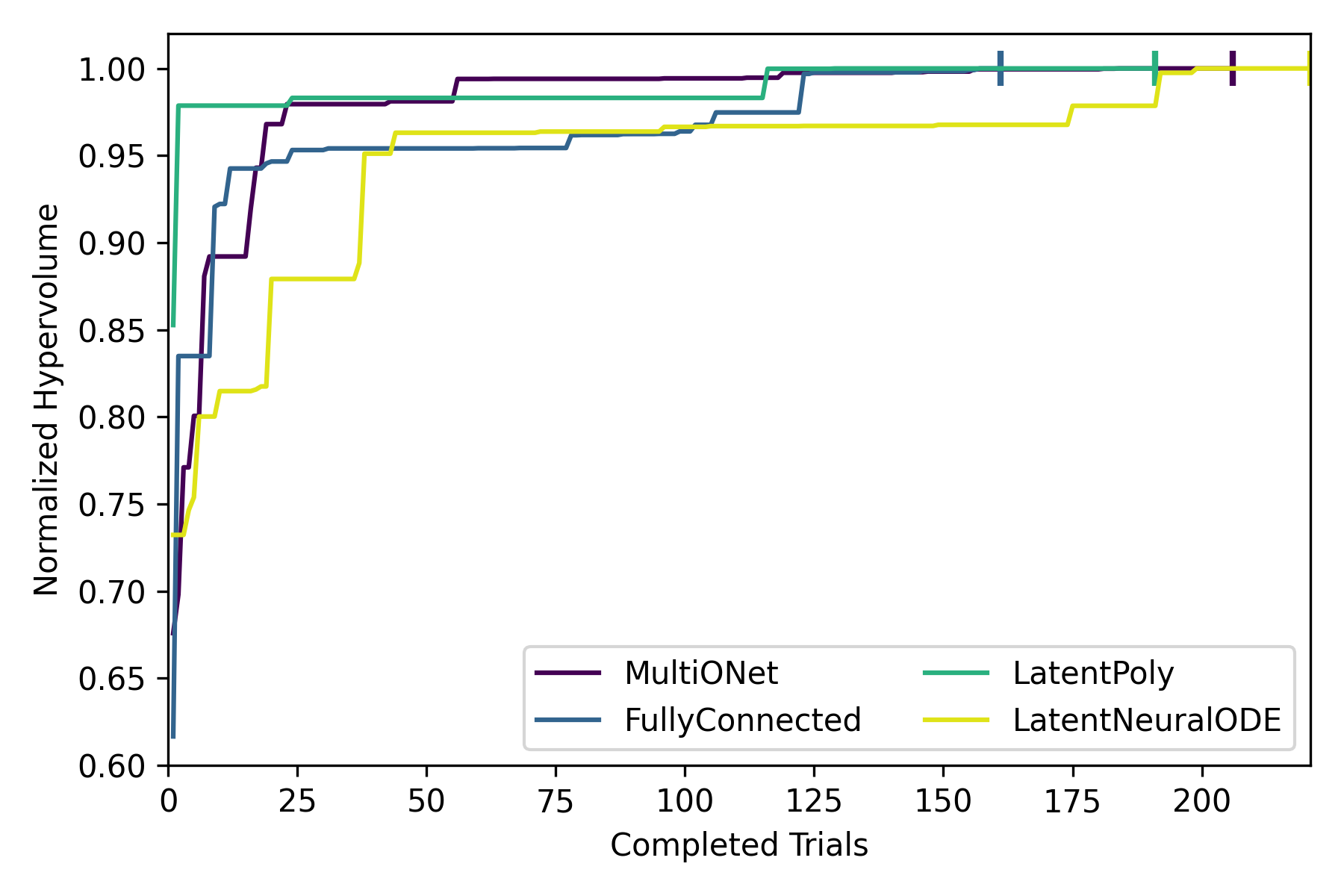}
\caption{Evolution of the normalized hypervolume spanned by the Pareto front during HPO for the primordial dataset. 
Vertical dashes indicate the final trial of each architecture-specific study. 
Most gains occur early in the optimization, followed by gradual saturation, suggesting diminishing returns from additional trials.}
  \label{fig:hypervolume}
\end{figure}

For each dataset and architecture, the optimization using the NSGA-II sampler yields a well-defined Pareto front. While the stochastic nature of the sampler does not allow us to make definitive statements about the state of convergence, we can use the development of the area covered by the Pareto front over the course of the tuning as a post-hoc heuristic to judge the progress of the optimization procedure. These areas were computed with respect to a reference point 10\% worse than the worst values observed across all trials in both objectives. Figure \ref{fig:hypervolume} shows the evolution of this area, normalized to the final Pareto-front area, over the course of the studies performed for the primordial dataset. Corresponding plots for the other datasets show similar behavior (see Appendix \ref{app:hyperparams:hypervolume}). We observe a saturation effect toward the end of the run, with most gains made early in the optimization procedure. This does not prove convergence, but indicates that substantial further gains are unlikely if the optimization were extended.

The resulting Pareto fronts exhibit substantial, architecture-dependent variability in both accuracy and efficiency, typically spanning factors of two to five with occasional extremes around an order of magnitude for both LAE$_{99}$ and inference time. The \texttt{FCNN} architecture displays the broadest accuracy spread, reflecting a strong sensitivity to hyperparameter choice, whereas the latent-evolution surrogates (\texttt{LP}, \texttt{LNODE}) show comparatively flat accuracy fronts but several-fold differences in runtime. 

Figure \ref{fig:pareto_fronts} exemplifies the resulting Pareto fronts for the primordial dataset, highlighting both the selected configuration (green) and the configuration with the lowest LAE$_{99}$ (red). These plots qualitatively illustrate that selecting the most accurate configuration without regard for efficiency often yields only a modest error reduction at the cost of a substantial increase in inference time. A quantitative analysis confirms this: in five out of the sixteen studies, the Pareto-selected configuration coincides with the lowest-error configuration. In the remaining eleven cases, prioritizing accuracy alone yields only modest error reductions at substantially increased inference cost. Averaged across these studies, LAE$_{99}$ decreases by $\sim$15\% at the expense of a $\sim$170\% increase in inference time. Examined on a per-study basis, this trade-off is highly asymmetric: reducing the error by 1\% typically requires an $\sim$20\% increase in inference time, with costs ranging from a few percent to more than 75\% depending on dataset and architecture. These results demonstrate that systematic, dual-objective tuning exposes a broad accuracy -- efficiency trade-off surface that would remain hidden under single-objective or manual adjustment. Even within the limited search budget used here, the optimizer consistently identifies configurations offering substantial efficiency gains for minimal accuracy loss, highlighting the practical value of explicit multi-objective optimization in surrogate selection.

\begin{figure*}[t]
  \centering
  \includegraphics[width=0.9\textwidth]{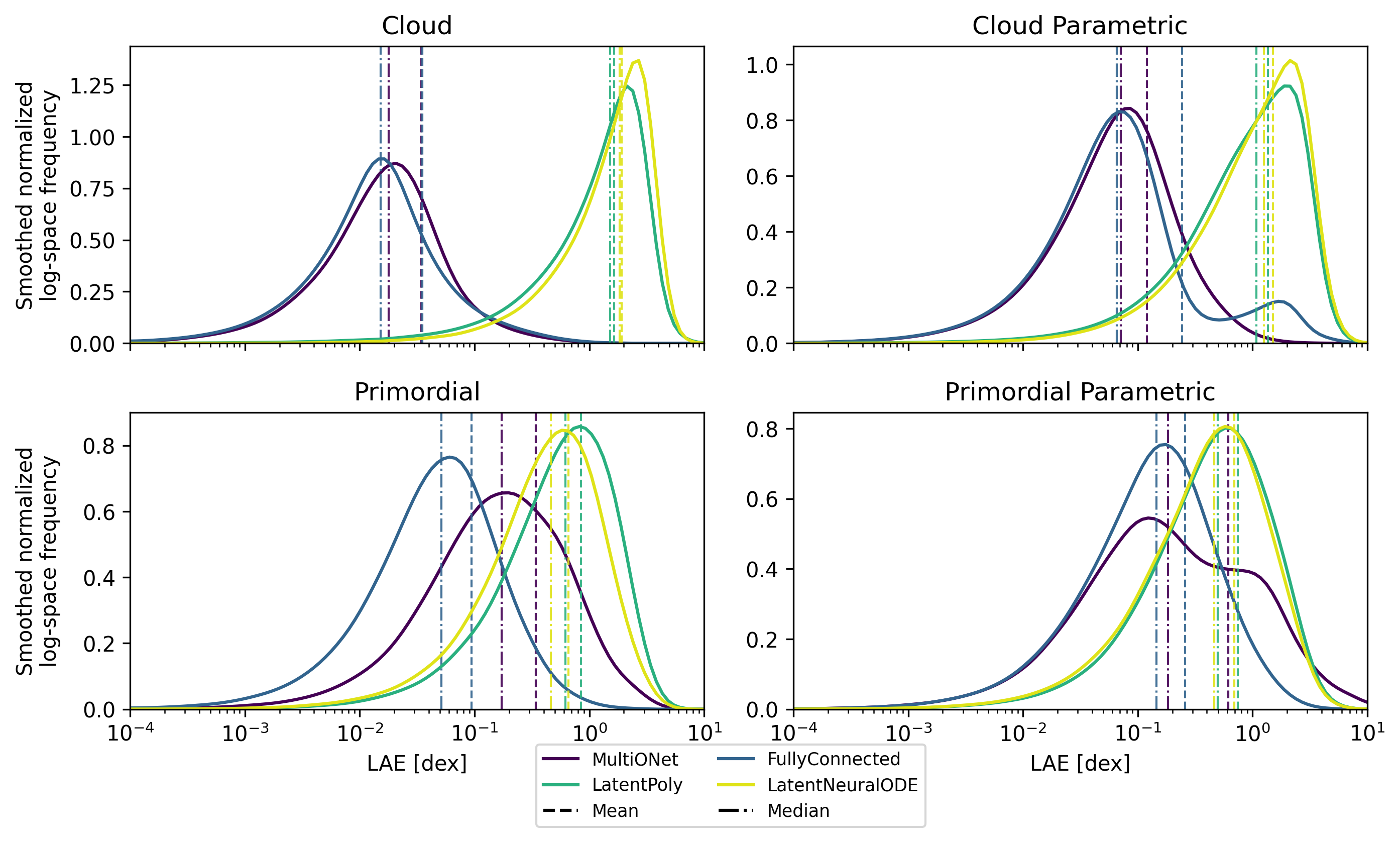}
\caption{Smoothed histograms of the LAE on the test set across surrogates and datasets, constructed in log-space and shown alongside the corresponding mean and median values. 
Latent-evolution surrogates exhibit systematically higher LAE values, with distribution peaks, means, and medians typically closely aligned. 
In contrast, for fully connected models the mLAE is frequently shifted toward higher values than the distribution peak, indicating a stronger influence of comparatively rare high-error predictions despite lower typical errors.}
  \label{fig:error_dists}
\end{figure*}

\subsection{Surrogate performance}
\label{subsec:accuracy}

An overview of selected performance metrics is provided in Table \ref{table:results}. Metrics directly related to prediction accuracy include the mLAE, the LAE$_{99}$, and the median relative error (mRE)\footnote{While we deem relative error metrics to be suboptimal, we still want to provide at least one relative error metric for reference and choose the median RE because it is less affected by a few strong outliers.}. For completeness, additional diagnostics on worst-case errors and temperature-only accuracy are reported in Appendix \ref{app:additional_accuracy}. Computational demand is quantified by the inference time, measured as the mean and standard deviation over five measurements of the time required to process the full test set on an NVIDIA TITAN Xp\footnote{
Inference is primarily compute-bound rather than memory-bound for the surrogate models considered here.
Peak GPU memory usage remained below 431\,MB across all experiments, making the NVIDIA TITAN Xp a convenient reference platform rather than a requirement.
}. These values should be interpreted only as a relative measure of computational demand, as absolute inference times depend strongly on hardware and on the size of the test set.

Overall, \texttt{FCNN} dominates these performance metrics. It consistently exhibits the lowest inference time  and lowest mRE across all datasets, often by a large margin, and achieves the lowest mLAE and LAE$_{99}$ on both primordial datasets. It is also nearly on par with \texttt{MON} on the cloud dataset, with cloud parametric being the only case where \texttt{MON} clearly outperforms it in mLAE and LAE$_{99}$. While the low inference time of \texttt{FCNN} is expected given its simple structure and efficient GPU execution, its consistently superior accuracy is noteworthy.

More generally, we observe a clear divide between two classes of architectures: the fully connected models (\texttt{FCNN}, \texttt{MON}) outperform the latent-evolution models (\texttt{LNODE}, \texttt{LP}) across all datasets, often by a wide margin. The former class consists of plain feed-forward networks that differ mainly in how they handle structural differences between inputs, and make few assumptions about the underlying data-generating process (low inductive bias). In contrast, the latter class features autoencoder-like bottlenecks and differs primarily in the choice of latent evolution mechanism, thereby imposing explicit assumptions about the dimensionality and structure of the latent dynamics (high inductive bias). The divide between these two classes is more pronounced on the cloud dataset variants than on the primordial dataset variants. Interestingly, the fully connected architectures generally perform better on the cloud variants than on the primordial variants across all three accuracy metrics\footnote{Comparisons are made separately for parametric and nonparametric variants.}, whereas the opposite trend is observed for the latent-evolution architectures. This suggests that the encoder–decoder structure or the assumed latent evolution may be ill suited to the cloud datasets. Key differences between the datasets that may contribute to this discrepancy are the larger number of species (38 versus 10), smaller average gradients, and the wider dynamic range of abundances in the cloud datasets.

A similar pattern emerges when comparing parametric and nonparametric variants: fully connected architectures achieve markedly higher accuracy on nonparametric datasets, whereas latent-evolution architectures perform similarly on both variants, occasionally even favoring the parametric case. The error distributions shown in Fig. \ref{fig:error_dists} complement the scalar metrics in Table \ref{table:results} by illustrating variability in model predictions across chemical states. They largely mirror the ranking observed in the summary statistics: fully connected architectures exhibit lower typical errors and tighter distributions, while latent-evolution architectures show broader spreads extending to higher error ranges, consistent with their larger LAE$_{99}$ values.

\begin{figure*}
  \centering
  \includegraphics[width=0.87\textwidth]{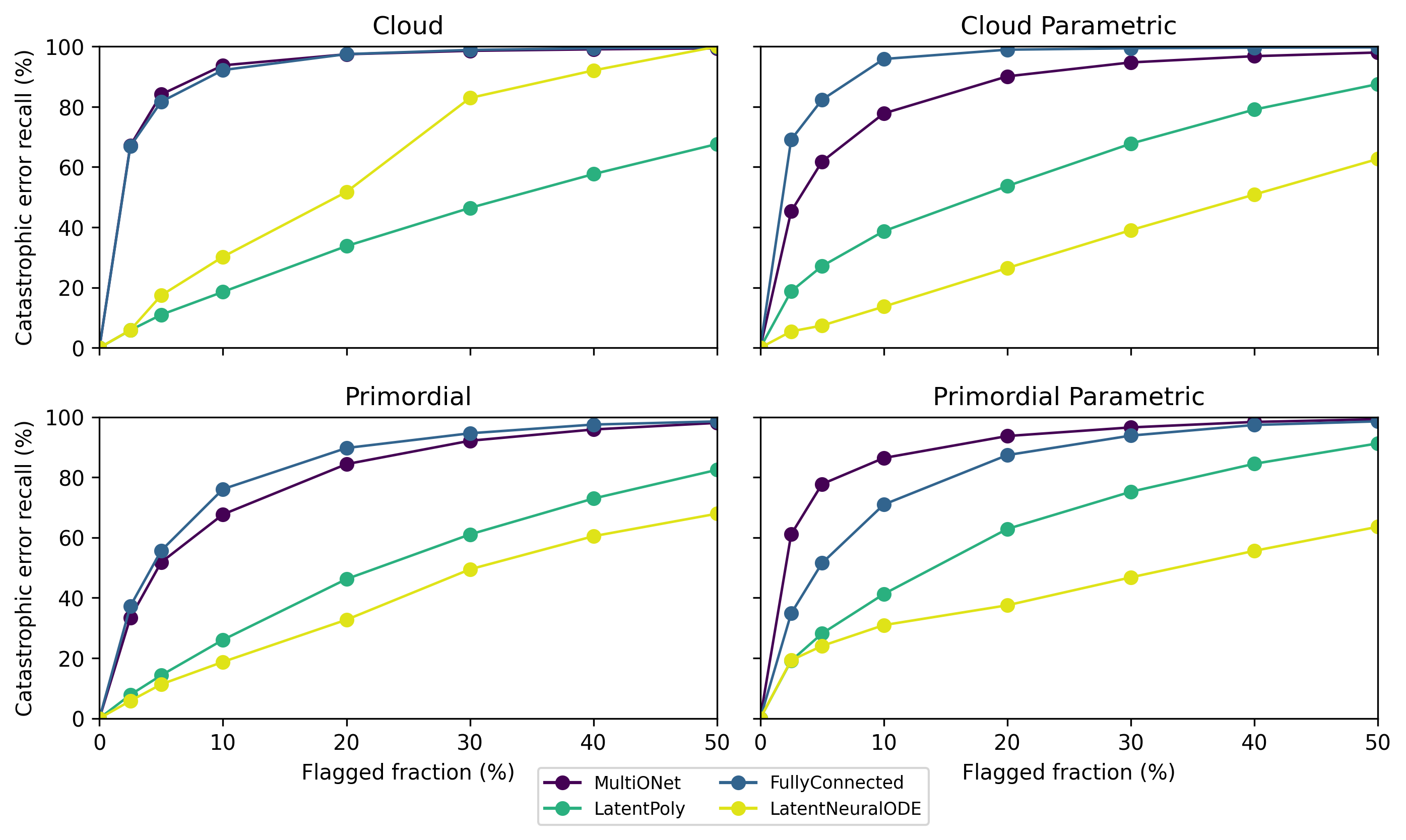}
  \caption{Recall of catastrophic errors on the test set as a function of the fraction of flagged predictions. Predictions are ranked by the predicted uncertainty (mLU) estimated from a DE ($M=5$), and increasing fractions of the most uncertain predictions are flagged. Catastrophic errors are defined as predictions whose log-space error exceeds the 99th percentile (LAE$_{99}$ in Table \ref{table:results}). Fully-connected surrogates achieve higher recall at lower flagged fractions than latent-evolution models, indicating more effective uncertainty-based error detection.}
  \label{fig:uq_detection}
\end{figure*}

With respect to the second optimization objective, inference time, we observe clear trends, albeit not as clearly grouped by architecture type. Within each dataset, \texttt{FCNN} is consistently the fastest model, whereas \texttt{LNODE} is by far the slowest, incurring nearly two orders of magnitude higher inference time in extreme cases. This reflects the high computational demand of the numerical integration required to evolve latent trajectories. In contrast, \texttt{LP} consistently outperforms \texttt{MON}, ranking second across datasets, indicating that an encoder–decoder structure does not necessarily incur high runtime overhead when the latent evolution is expressed in closed form rather than via numerical integration.

Interestingly, computational efficiency shows no direct correspondence to model size. For example, \texttt{FCNN} contains up to an order of magnitude more parameters than \texttt{LP} in some cases, yet remains faster by a factor of two to three. Even the architecturally much more similar \texttt{MON} is slower than \texttt{FCNN} at lower parameter count. This suggests that increased computational cost arises from non-standard operations -- the scalar product in \texttt{MON}, polynomial evaluation in \texttt{LP}, and, most prominently, numerical integration in \texttt{LNODE} -- whose forward passes cannot fully exploit highly optimized compute kernels.

\subsection{Uncertainty quantification}
\label{subsec:UQ}

We first investigate whether a DE achieves improved accuracy compared to a single model. This is assessed by comparing the mLAE of the single model with mLAE (DE) in Table \ref{table:results}, which is the mLAE obtained using the ensemble mean as prediction, rather than the prediction of the main model. For clarity, we note that CODES forms the DE by training $M-1$ additional models (with $M=5$ in our case), incorporating the main model in the DE. This comparison reveals substantial improvements in mLAE over the single-model case for \texttt{MON} across all datasets, as well as for \texttt{FCNN} on the parametric dataset variants. On the nonparametric variants, the \texttt{FCNN} DE exhibits slightly higher mLAE than the single model. For the latent-evolution surrogates, only minor changes in mLAE are observed, without a consistent trend in either direction. One possible explanation is that models within latent-evolution ensembles tend to agree more closely with each other than those in fully connected ensembles. This is reflected in the mean log uncertainty (mLU), computed as the mean standard deviation of ensemble predictions across the test set. The mLU is of similar magnitude across surrogate classes, but it must be interpreted relative to the corresponding ensemble mLAE values. After all, the size of an error bar is only representative in relation to the magnitude of the value it refers to. Thus, on average, predictions within latent-evolution ensembles exhibit a much smaller spread relative to the true error magnitude.

To disentangle uncertainty calibration from absolute scale, we additionally report the UQ Pearson correlation coefficient (UQ PCC) between the ensemble standard deviation and the corresponding LAE across all test-set predictions, i.e., between predicted uncertainty and actual error. By nature, the PCC only measures linear correlation, remaining agnostic to the scaling of the quantities. The PCC differs systematically between surrogate classes: it exceeds 0.5 in all cases for fully connected models, often by a large margin, but remains below 0.5 for all latent-evolution models. The smallest correlations are observed for the cloud dataset variants, where the correlation is close to 0 for the latter surrogate class. In summary, these metrics indicate that DEs based on fully connected models more accurately estimate the magnitude of their prediction error than those based on latent-evolution models.

Finally, we investigate whether DEs can realistically trigger a fallback to the numerical solver in cases of catastrophic prediction failure. For the purposes of this investigation, we define errors that exceed the 99th percentile as catastrophic errors. This means that for each surrogate-dataset pair, catastrophic errors are those above LAE$_{99}$ in Table \ref{table:results}. 
Using this definition, Fig. \ref{fig:uq_detection} shows how reliably catastrophic errors can be identified based on the predicted uncertainty.
Predictions are ordered by their mLU, and increasing fractions of predictions are flagged as unreliable (from most to least uncertain).
For each such fraction, we report the corresponding recall (true positive rate) of catastrophic errors, i.e., the fraction of predictions with errors exceeding LAE$_{99}$ that are correctly flagged.
Plotting recall against the fraction of flagged predictions makes the trade-off between detection performance and fallback cost explicit, without relying on a specific uncertainty threshold.

In an application setting, the approach would be inverted: we would set a threshold value based on the desired recall or the affordable loss of efficiency. This efficiency loss is an inherent consequence of UQ-based fallback mechanisms, since the numerical solver may be orders of magnitude slower. In Fig. \ref{fig:uq_detection} we observe that the fully connected ensembles capture their prediction errors much more reliably than latent-evolution ensembles. While the former detect more than 80\% of prediction errors at a threshold that flags less than 20\% of predictions in all cases, the latter only reach this recall at much higher fractions. This aligns with the observations about the UQ PCC -- fully connected ensembles predict the magnitude of their prediction error more reliably, translating directly to higher catastrophic error detection efficiency. This behavior is robust to the precise definition of catastrophic error: repeating the analysis with a lower threshold yields qualitatively identical recall trends across architectures (see Appendix \ref{app:error_detection}).

\subsection{Error propagation}
\label{subsec:error_prop}

\begin{figure*}[t]
  \centering
  \includegraphics[width=0.87\textwidth]{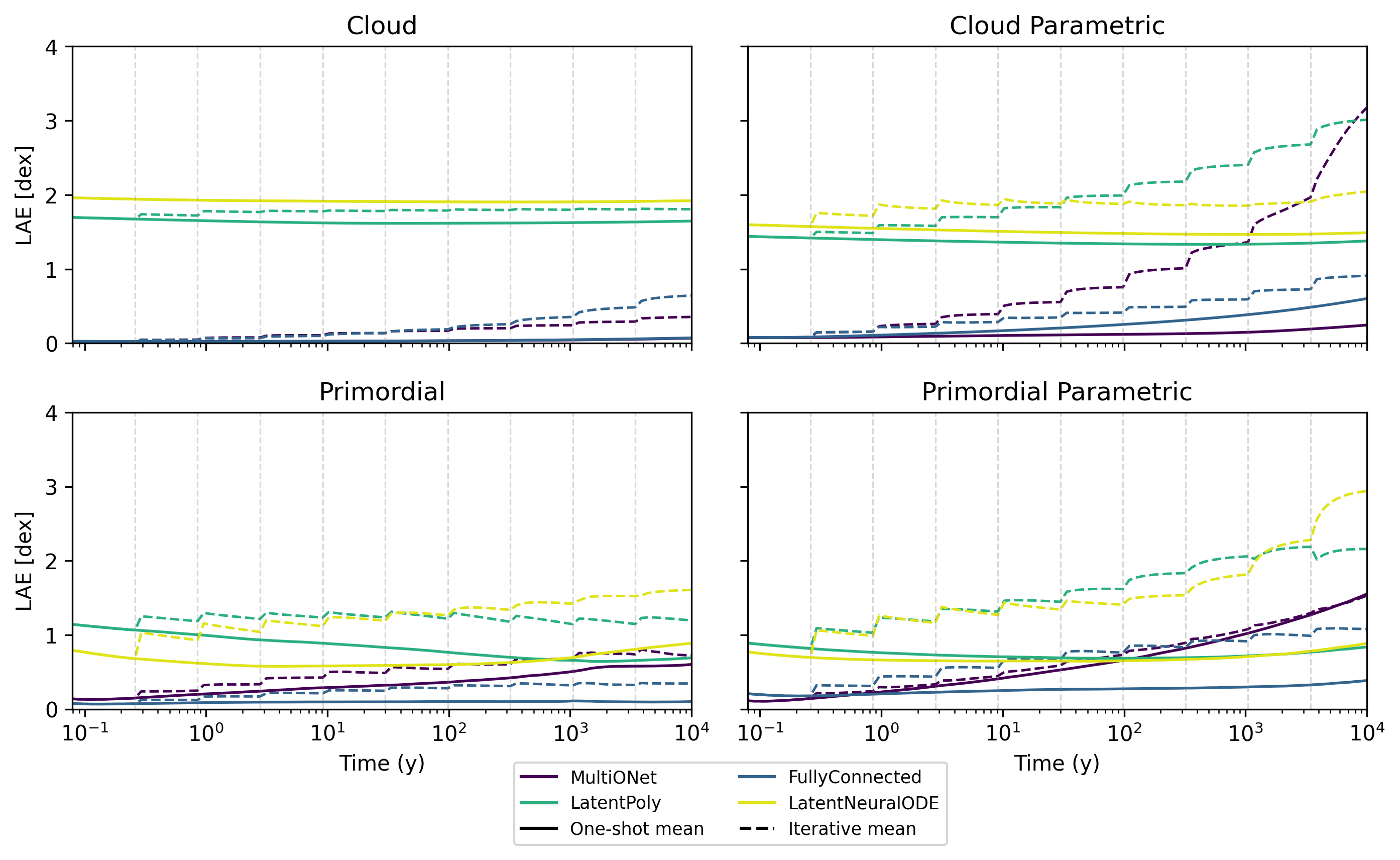}
  \caption{Evolution of the mLAE on the test set over time across surrogates and datasets. To investigate error accumulation over time, we compared the one-shot mean, where predictions are obtained by providing the initial state and the desired time step, with the iterative mean. For the latter, we obtained predictions by dividing the domain into intervals of $i=10$ time steps (indicated by the vertical dashed lines), and using the predicted abundances at the end of one interval as initial state for the next interval. In this iterative setting, fully connected surrogates generally exhibit a more pronounced accrual of error over time, while latent-evolution models prove more robust.}
  \label{fig:iterative_errors}
\end{figure*}

Since surrogate models are intended to replace costly numerical solver calls at every hydrodynamical time step, their stability under repeated application is a key property. In a full simulation, abundances would be modified by hydrodynamics between surrogate calls. As a proxy, we instead reuse the surrogate’s prediction over one part of the time domain as the initial state for the next, following \citet{pelle25}. In CODES, the time domain is divided into segments of $i$ time steps, and predictions are generated iteratively by using the final state of each segment as the initial condition for the next. Figure \ref{fig:iterative_errors} compares the LAEs across datasets and architectures for $i=10$ resulting from both iterative prediction and the standard one-shot prediction across the entire time domain (i.e., predicting all time steps directly from the initial state). Additionally, Table \ref{table:results} reports the mean iterative log-space absolute error (mLAE$_\mathrm{iter}$), corresponding to the mean of the trajectories shown in Fig. \ref{fig:iterative_errors}.   In all cases, the mean iterative error exceeds the one-shot error, indicating that repeated surrogate application degrades accuracy. Additional tests show that the degree of error accumulation depends on the choice of $i$, with more frequent surrogate calls leading to larger accumulated errors (see Appendix \ref{app:iterative_eval} for $i=3$).

The two architectural classes again exhibit distinct behavior in the iterative setting, with higher-inductive-bias models showing greater robustness. Fully connected surrogates, despite their superior one-shot accuracy across most datasets, accumulate errors more rapidly under iterative application than latent-evolution architectures. This is most apparent on the cloud dataset, where the latter class of surrogates incurs almost no loss in accuracy when evaluated iteratively. 
Although latent-evolution models typically start with higher errors at early times, the effect is strong enough that \texttt{MON}, despite achieving the lowest initial error on the cloud parametric dataset, exhibits the highest error at the end of the time domain. These findings suggest that architectures with stronger inductive bias, while underperforming in standard accuracy metrics, capture more of the governing dynamics and are therefore more reliable when making predictions for previously unseen initial conditions.

\section{Discussion}    
\label{sec:discussion} 

The results presented in Sect. \ref{sec:results} offer a coherent picture of how the proposed surrogate architectures behave when applied to realistic, challenging astrochemical datasets. We identify multiple key takeaways from these investigations.

\paragraph{Hyperparameter tuning is crucial.} As evidenced by the substantial span of the Pareto fronts across both optimization objectives, chosen hyperparameters strongly influence surrogate performance across architectures. The interdependence of hyperparameters and the complexity of the search space render manual optimization impractical, necessitating systematic tuning. Most importantly, however, our findings in Sect. \ref{subsec:hyperparam_results} emphasize the importance of making the desired surrogate properties explicit optimization objectives. The observed trade-off -- where in some cases a fivefold increase in inference time yielded only minimal accuracy gains -- would not be visible if efficiency were excluded from the optimization procedure. Moreover, this trade-off emerges using a sampling algorithm explicitly designed to explore the Pareto front and identify configurations that balance both objectives. A single-objective optimization using a corresponding algorithm (e.g., the Tree-Parzen Estimator) may optimize more aggressively toward accuracy, sacrificing even more efficiency in the process. 

\vspace{-0.7em}

\paragraph{Inductive bias influences performance.}

Across Sect. \ref{sec:results}, we observe pronounced differences between two classes of architectures: fully connected surrogates and latent-evolution surrogates. These groups differ systematically in the amount of inductive bias they introduce. The former make few assumptions about the underlying problem, whereas the latter are explicitly designed for trajectories arising from coupled ODEs. On accuracy alone, the fully connected architectures are superior in many cases, especially \texttt{FCNN}. While we do not claim a definitive explanation, we hypothesize that for the data considered here -- complex coupled ODEs with additional external parameters -- strong inductive biases, such as enforced low-dimensional latent trajectories, may restrict model flexibility more than they help regularize learning. This interpretation is consistent with observations reported by \cite{grassi:2021}, where a latent-evolution model was found to effectively time-average latent dynamics, thereby attenuating high-frequency components associated with stiff behavior. For \texttt{LNODE} in particular, an additional contributing factor may be vanishing gradients, which pose a fundamental challenge for neural differential equations acting on stiff problems \citep{fronk_vanishing_2025}.
The picture is more nuanced for inference time, which largely depends on the computational cost of the non-standard operations involved. However, similar distinctions between the two groups emerge in UQ, where fully connected architectures are more reliable, and in iterative evaluation, where latent-evolution models prove more robust. Both observations are consistent with differences in inductive bias: stronger bias reduces variability among ensemble members, degrading uncertainty estimates, while simultaneously increasing robustness under challenging conditions and limiting performance degradation under iterative predictions.

\vspace{-1.5em}

\paragraph{Deep ensembles provide reliable uncertainty estimates.}

The investigations in Sect. \ref{subsec:UQ} show that DEs with $M=5$ members provide a reliable mechanism for detecting catastrophic prediction errors. We observe pronounced differences in detection efficiency between architectures: latent-evolution models achieve comparable recall only when flagging a much larger fraction of predictions. The fully connected surrogates, however, detect catastrophic errors with high efficiency. This makes a confidence-triggered fallback mechanism to the numerical solver, which we expect to improve overall accuracy and significantly enhance trajectory stability, a realistic option.

\paragraph{Performance depends on representative data.}

Directly investigating trajectory stability via iterative surrogate calls in Sect. \ref{subsec:error_prop}, we again find pronounced differences between architectures.
The latent-evolution architectures prove more robust, whereas fully connected models exhibit pronounced error accrual. While strategies such as adding noise during training can mitigate error accumulation \citep{holdship:2021}, we conjecture that the dominant cause is initial-condition drift: repeated surrogate evaluations expose the model to states outside its training domain. A perfect surrogate that fully represents the underlying ODE would not be affected by this drift. However, for data-driven surrogates, leaving the training domain degrades accuracy, as evidenced by the jumps in mean error at the start of each iteration interval. This hypothesis is consistent with the observed robustness of latent-evolution architectures--their higher inductive bias promotes learning the underlying mathematical structure rather than merely reproducing observed trajectories.

Regardless of the architecture of choice, this observation emphasizes the importance of training data being representative of the later application. For surrogate deployment in large-scale astrophysical simulations, this implies that the distribution of initial abundances in the training data should closely match that encountered in solver calls during the simulation. While this distribution is a priori unknown, a practical approach is to approximate it using small-scale simulations driven by the accurate but expensive numerical solver.

\paragraph{Limitations}

In this work, we compare surrogates across multiple performance dimensions, four datasets, and four architectures. While CODES provides a rigorous framework for this comparison, ensuring that evaluations are fully balanced and representative remains challenging due to architectural differences and application-specific implementation details.
One example is the measurement of inference time. We measure the time required to predict all trajectories in the test set using the same batched setup as during training, which favors architectures that benefit from batching. Additionally, for \texttt{LNODE}, we opt for evaluating entire trajectories at once, allowing a single numerical integration, rather than letting the solver solve anew for each time step. Whether these choices fully reflect realistic applications remains open.

A natural question is how these surrogate runtimes compare to those of full chemical solvers.  We do not attempt such a comparison here for multiple reasons. First, establishing a fair baseline is non-trivial: astrochemical solvers vary widely in implementation details, may run on CPU or GPU, and performance is tightly coupled to integrator settings.
Additionally, as described in Sect. \ref{sec:intro}, neural surrogates map naturally to modern GPU hardware, where dense tensor operations dominate performance. Their runtime characteristics therefore differ fundamentally from those of ODE solvers, making direct one-to-one comparisons difficult to interpret.
Furthermore, the literature already provides multiple demonstrations that neural surrogates outperform traditional solvers by orders of magnitude. \cite{silke24} find a speed-up by a factor of 24-28, while \cite{branca24} measure a factor of 128, and \cite{sulzer:2023} observed a speed-up of factor 55 for their original implementation of \texttt{LNODE} and an impressive factor 4270 for the original \texttt{LP}.

While we performed thorough relative comparisons, assessing surrogate accuracy in absolute terms remains challenging, primarily because the impact of a given error on a downstream simulation is difficult to determine. In the future, integrating surrogates into fully differentiable magnetohydrodynamic frameworks such as \texttt{astronomix} \citep{storcks_astronomix_2025} may enable new forms of sensitivity analysis, allowing errors to be traced back to individual species and predictions, which is not possible in a benchmark decoupled from the simulation. We also do not explicitly enforce mass or charge conservation, and leave conservation-aware training or renormalization strategies for future work. In addition to this, we did not compare our accuracy results to other works for two reasons: we use different metrics to measure accuracy (see Sect. \ref{subsec:metrics}), and we work on different datasets. 

To facilitate future comparisons, our datasets are publicly available and CODES automates optimization, training, and evaluation on these and other datasets. Further work is needed to quantify dataset difficulty, potentially via measures such as stiffness (as a proxy for dynamical complexity) or the hypervolume of the initial-condition space.

Another limitation concerns details of the HPO setup. While we consider our choices well motivated, aspects such as accuracy metrics, trial budgets, and optimizer selection remain largely heuristic. We identify the systematic exploration of these HPO design choices in the context of astrochemical surrogate modeling as a key opportunity for future research.

A limitation of our UQ analysis is that it relies on the test split of the same Sobol-sampled dataset used for training and validation. While the initial conditions in this split are unseen during training and optimization, they stem from the same underlying distribution. Although our results suggest that ensemble spread should increase for out-of-distribution inputs, further investigation is required to rule out failure modes such as mode collapse, where ensemble uncertainty may not reliably reflect prediction error.

Finally, we would like to emphasize that this work does not claim to present optimal or best-performing surrogates. While to our knowledge, the architectures presented here were optimized extensively for ideal performance on each dataset, we are aware that, as ever in machine learning, performance depends substantially on finding the optimal configuration of an array of training tools and architectural details. We do not expect substantial gains from retuning hyperparameters alone, as this is precisely what the HPO procedure addresses. But it is entirely possible that changes to architectures or training, for example adding a new loss term, altering the optimizer, or modifying regularization, could significantly improve performance. 

Similarly, we do not claim that the architectures we present here are ideal. Our aim is to encourage architecture research by providing a systematic and rigorous approach to measuring and comparing performance across architectures. To facilitate this, we proposed challenging datasets that capture the complexities of the application context, a dual-objective optimization procedure to ensure surrogates perform as intended, log-space metrics to accurately evaluate surrogate performance, and additional investigations of surrogate behavior. These proposals are bundled in the open-source benchmarking suite CODES, which is well-documented and easily extensible. 

\paragraph{Surrogates are viable for realistic astrochemical dynamics.}

Despite these limitations, the observed accuracy gains, substantial improvements in computational efficiency, and the additional safety provided by DE uncertainty estimation suggest that replacing numerical solvers with surrogates in complex astrophysical simulations is generally feasible. Our analysis indicates that, for the datasets considered here, fully connected models -- most notably \texttt{FCNN} -- are better suited to this task: they are more accurate, faster, and provide more reliable uncertainty estimates. It is, however, plausible that other datasets may benefit from the additional regularization and physical structure imposed by latent-evolution architectures.
CODES facilitates informed, data-driven choices of surrogate architecture and configuration for a given task.

\section{Conclusions}
\label{sec:conclusion}

In this work, we proposed a systematic approach to identifying well-performing surrogate models for a given task, as represented by a dataset. This was achieved by separately optimizing four architectures for accuracy and inference time, followed by a multidimensional comparison of their behavior. For this comparison, we introduced log-space accuracy metrics tailored to the large dynamic range characteristic of astrochemical data. The entire pipeline--tuning, training, and evaluation--is implemented in the CODES benchmarking framework. 

We applied this procedure to four different datasets that represent realistic application scenarios, characterized by large dynamic range, relative stiffness, and additional challenges such as external ODE parameters. Our findings across surrogates and datasets are multifaceted: joint optimization for accuracy and inference time reveals an intrinsic trade-off between these objectives. We found that across accuracy, UQ, and error propagation measurements, the architectures fall into two groups. This grouping aligns closely with the degree of inductive bias incorporated by each architecture. For the datasets studied here, higher inductive bias -- in the form of explicit latent-space time evolution -- improves robustness in challenging settings but comes at the cost of reduced accuracy and less reliable DE uncertainty estimates.

A natural next step is to deploy the optimized surrogates within full-scale astrophysical simulations. This will reveal whether the performance characteristics observed in benchmark studies -- accuracy, UQ with fallback mechanisms, and error propagation -- translate to practical application settings. We expect this transition to a real application setting to introduce new challenges, while also enabling deeper insight into practical surrogate behavior. To support and encourage such efforts, CODES is made available as an open framework, and we welcome its use and extension by the community to explore additional datasets, architectures, and application scenarios.

\section*{Code and data availability}

Our benchmarking framework \textsc{CODES} is publicly available on GitHub at
\href{https://github.com/AstroAI-Lab/CODES-Benchmark}{https://github.com/AstroAI-Lab/CODES-Benchmark}.
A dedicated documentation website including a tutorial and auto-generated API documentation is provided at
\href{https://astroai-lab.de/CODES-Benchmark}{https://astroai-lab.de/CODES-Benchmark}. The datasets used in this work are publicly available on Zenodo under DOI \href{https://doi.org/10.5281/zenodo.18018572}{10.5281/zenodo.18018572}.
Additional datasets included in \textsc{CODES} are also publicly available via Zenodo and are automatically downloaded on demand.

\begin{acknowledgements}
This work was funded by the Carl-Zeiss-Stiftung through the NEXUS program and supported by the Deutsche Forschungsgemeinschaft (DFG, German Research Foundation) under Germany’s Excellence Strategy EXC~2181/1--390900948 (the Heidelberg STRUCTURES Excellence Cluster). The authors acknowledge support from the IWR Collaborate Program at Heidelberg University. We also acknowledge the usage of the AI clusters Tom and Jerry, funded by Field of Focus~2 of Heidelberg University. 

This work benefited from discussions at the Lorentz Center workshop “Machine Learning in Astrochemistry”. In particular, we thank Immanuel Sulzer, Simon Glover, Felix Priestley, and Tommaso Grassi for valuable exchanges.
\end{acknowledgements}

\bibliographystyle{bib/aa}
\bibliography{bib/main}     

@article{galli:1998,
  adsurl = {https://ui.adsabs.harvard.edu/abs/1998A&A...335..403G},
  archiveprefix = {arXiv},
  author = {{Galli}, Daniele and {Palla}, Francesco},
  doi = {10.48550/arXiv.astro-ph/9803315},
  eprint = {astro-ph/9803315},
  journal = {\aap},
  month = {Jul},
  pages = {403-420},
  primaryclass = {astro-ph},
  title = {{The chemistry of the early Universe}},
  volume = {335},
  year = {1998}
}

@article{glover:2008,
  adsurl = {https://ui.adsabs.harvard.edu/abs/2008MNRAS.388.1627G},
  archiveprefix = {arXiv},
  author = {{Glover}, S.~C.~O. and {Abel}, T.},
  doi = {10.1111/j.1365-2966.2008.13224.x},
  eprint = {0803.1768},
  journal = {\mnras},
  month = {Aug},
  number = {4},
  pages = {1627-1651},
  primaryclass = {astro-ph},
  title = {{Uncertainties in H$_{2}$ and HD chemistry and cooling and their role in early structure formation}},
  volume = {388},
  year = {2008}
}

@article{caselli:2012,
  adsurl = {https://ui.adsabs.harvard.edu/abs/2012A&ARv..20...56C},
  archiveprefix = {arXiv},
  author = {{Caselli}, Paola and {Ceccarelli}, Cecilia},
  doi = {10.1007/s00159-012-0056-x},
  eid = {56},
  eprint = {1210.6368},
  journal = {\aapr},
  month = {Oct},
  pages = {56},
  primaryclass = {astro-ph.GA},
  title = {{Our astrochemical heritage}},
  volume = {20},
  year = {2012}
}

@article{pallottini:2017_b,
  adsurl = {http://adsabs.harvard.edu/abs/2017MNRAS.471.4128P},
  archiveprefix = {arXiv},
  author = {{Pallottini}, A. and {Ferrara}, A. and {Bovino}, S. and
{Vallini}, L. and {Gallerani}, S. and {Maiolino}, R. and
{Salvadori}, S.},
  doi = {10.1093/mnras/stx1792},
  eprint = {1707.04259},
  journal = {\mnras},
  month = {November},
  pages = {4128-4143},
  title = {{The impact of chemistry on the structure of high-z
galaxies}},
  url = {https://ui.adsabs.harvard.edu/abs/2017MNRAS.471.4128P},
  volume = {471},
  year = {2017}
}

@article{lupi:2019,
  adsurl = {https://ui.adsabs.harvard.edu/abs/2019MNRAS.484.1687L},
  archiveprefix = {arXiv},
  author = {{Lupi}, Alessandro},
  doi = {10.1093/mnras/stz100},
  eprint = {1808.10184},
  journal = {\mnras},
  month = {Apr},
  number = {2},
  pages = {1687-1701},
  primaryclass = {astro-ph.GA},
  title = {{H$\_{2}$ chemistry in galaxy simulations: an improved
supernova feedback model}},
  url = {https://ui.adsabs.harvard.edu/abs/2019MNRAS.484.1687L},
  volume = {484},
  year = {2019}
}

@article{decataldo:2019,
  adsurl = {https://ui.adsabs.harvard.edu/abs/2019MNRAS.487.3377D},
  archiveprefix = {arXiv},
  author = {{Decataldo}, D. and {Pallottini}, A. and {Ferrara}, A. and {Vallini}, L. and {Gallerani}, S.},
  doi = {10.1093/mnras/stz1527},
  eprint = {2009.07860},
  journal = {\mnras},
  month = {Aug},
  number = {3},
  pages = {3377-3391},
  primaryclass = {astro-ph.GA},
  title = {{Photoevaporation of Jeans-unstable molecular clumps}},
  volume = {487},
  year = {2019}
}

@article{kim:2018,
  adsurl = {https://ui.adsabs.harvard.edu/abs/2018ApJ...859...68K},
  archiveprefix = {arXiv},
  author = {{Kim}, Jeong-Gyu and {Kim}, Woong-Tae and {Ostriker}, Eve C.},
  doi = {10.3847/1538-4357/aabe27},
  eid = {68},
  eprint = {1804.04664},
  journal = {\apj},
  month = {May},
  number = {1},
  pages = {68},
  primaryclass = {astro-ph.GA},
  title = {{Modeling UV Radiation Feedback from Massive Stars. II. Dispersal of Star-forming Giant Molecular Clouds by Photoionization and Radiation Pressure}},
  volume = {859},
  year = {2018}
}

@article{bovino19,
  adsurl = {https://ui.adsabs.harvard.edu/abs/2019BAAA...61..274B},
  author = {{Bovino}, S. and {Schleicher}, D.~R.~G. and {Grassi}, T.},
  journal = {Boletin de la Asociacion Argentina de Astronomia La Plata Argentina},
  month = {Aug},
  pages = {274-276},
  title = {{Computational Astrochemistry: importance, pitfalls and applications}},
  volume = {61},
  year = {2019}
}

@article{ferland:2017,
  adsurl = {https://ui.adsabs.harvard.edu/abs/2017RMxAA..53..385F},
  archiveprefix = {arXiv},
  author = {{Ferland}, G.~J. and {Chatzikos}, M. and {Guzm{\'a}n}, F. and {Lykins}, M.~L. and {van Hoof}, P.~A.~M. and {Williams}, R.~J.~R. and {Abel}, N.~P. and {Badnell}, N.~R. and {Keenan}, F.~P. and {Porter}, R.~L. and {Stancil}, P.~C.},
  doi = {10.48550/arXiv.1705.10877},
  eprint = {1705.10877},
  journal = {\rmxaa},
  month = {Oct},
  pages = {385-438},
  primaryclass = {astro-ph.GA},
  title = {{The 2017 Release Cloudy}},
  volume = {53},
  year = {2017}
}

@article{holdship:2017,
  adsurl = {https://ui.adsabs.harvard.edu/abs/2017AJ....154...38H},
  archiveprefix = {arXiv},
  author = {{Holdship}, J. and {Viti}, S. and {Jim{\'e}nez-Serra}, I. and {Makrymallis}, A. and {Priestley}, F.},
  doi = {10.3847/1538-3881/aa773f},
  eid = {38},
  eprint = {1705.10677},
  journal = {\aj},
  month = {Jul},
  number = {1},
  pages = {38},
  primaryclass = {astro-ph.GA},
  title = {{UCLCHEM: A Gas-grain Chemical Code for Clouds, Cores, and C-Shocks}},
  volume = {154},
  year = {2017}
}

@article{gray:2019,
  adsurl = {https://ui.adsabs.harvard.edu/abs/2019ApJ...887..161G},
  archiveprefix = {arXiv},
  author = {{Gray}, William J. and {Oey}, M.~S. and {Silich}, Sergiy and {Scannapieco}, Evan},
  doi = {10.3847/1538-4357/ab510d},
  eid = {161},
  eprint = {1910.12882},
  journal = {\apj},
  month = {Dec},
  number = {2},
  pages = {161},
  primaryclass = {astro-ph.GA},
  title = {{Catastrophic Cooling in Superwinds: Line Emission and Non-equilibrium Ionization}},
  volume = {887},
  year = {2019}
}

@article{olsen:2018,
  adsurl = {https://ui.adsabs.harvard.edu/abs/2018Galax...6..100O},
  archiveprefix = {arXiv},
  author = {{Olsen}, Karen P. and {Pallottini}, Andrea and {Wofford}, Aida and {Chatzikos}, Marios and {Revalski}, Mitchell and {Guzm{\'a}n}, Francisco and {Popping}, Gerg{\"o} and {V{\'a}zquez-Semadeni}, Enrique and {Magdis}, Georgios E. and {Richardson}, Mark L.~A. and {Hirschmann}, Michaela and {Gray}, William J.},
  doi = {10.3390/galaxies6040100},
  eid = {100},
  eprint = {1808.08251},
  journal = {Galaxies},
  month = {Sep},
  number = {4},
  pages = {100},
  primaryclass = {astro-ph.GA},
  title = {{Challenges and Techniques for Simulating Line Emission}},
  volume = {6},
  year = {2018}
}

@article{vallini:2018,
  adsurl = {https://ui.adsabs.harvard.edu/abs/2018MNRAS.473..271V},
  archiveprefix = {arXiv},
  author = {{Vallini}, L. and {Pallottini}, A. and {Ferrara}, A. and {Gallerani}, S. and {Sobacchi}, E. and {Behrens}, C.},
  doi = {10.1093/mnras/stx2376},
  eprint = {1709.03993},
  journal = {\mnras},
  month = {Jan},
  number = {1},
  pages = {271-285},
  primaryclass = {astro-ph.GA},
  title = {{CO line emission from galaxies in the Epoch of Reionization}},
  volume = {473},
  year = {2018}
}

@article{danehkar:2022,
  adsurl = {https://ui.adsabs.harvard.edu/abs/2022ApJ...937...68D},
  archiveprefix = {arXiv},
  author = {{Danehkar}, A. and {Oey}, M.~S. and {Gray}, W.~J.},
  doi = {10.3847/1538-4357/ac8cec},
  eid = {68},
  eprint = {2208.12030},
  journal = {\apj},
  month = {Oct},
  number = {2},
  pages = {68},
  primaryclass = {astro-ph.GA},
  title = {{Catastrophic Cooling in Superwinds. III. Nonequilibrium Photoionization}},
  volume = {937},
  year = {2022}
}

@article{kumar:2013MNRAS,
  adsurl = {https://ui.adsabs.harvard.edu/abs/2013MNRAS.431..455K},
  archiveprefix = {arXiv},
  author = {{Kumar}, A. and {Fisher}, R.~T.},
  doi = {10.1093/mnras/stt171},
  eprint = {1302.0330},
  journal = {\mnras},
  month = {May},
  number = {1},
  pages = {455-476},
  primaryclass = {astro-ph.SR},
  title = {{The astrochemical evolution of turbulent giant molecular clouds: physical processes and method of solution for hydrodynamic, embedded starless clouds}},
  volume = {431},
  year = {2013}
}

@article{grassi:2014,
  adsurl = {http://adsabs.harvard.edu/abs/2014MNRAS.439.2386G},
  author = {{Grassi}, T. and {Bovino}, S. and {Schleicher}, D.~R.~G. and 
{Prieto}, J. and {Seifried}, D. and {Simoncini}, E. and {Gianturco}, F.~A.
},
  doi = {10.1093/mnras/stu114},
  journal = {\mnras},
  month = {Apr},
  pages = {2386-2419},
  title = {{KROME - a package to embed chemistry in astrophysical simulations}},
  volume = {439},
  year = {2014}
}

@article{ziegler:2016,
  adsurl = {https://ui.adsabs.harvard.edu/abs/2016A&A...586A..82Z},
  author = {{Ziegler}, U.},
  doi = {10.1051/0004-6361/201527262},
  eid = {A82},
  journal = {\aap},
  month = {Feb},
  pages = {A82},
  title = {{A chemical reaction network solver for the astrophysics code NIRVANA}},
  volume = {586},
  year = {2016}
}

@article{smith:2017,
  adsurl = {https://ui.adsabs.harvard.edu/abs/2017MNRAS.466.2217S},
  archiveprefix = {arXiv},
  author = {{Smith}, Britton D. and {Bryan}, Greg L. and {Glover}, Simon C.~O. and {Goldbaum}, Nathan J. and {Turk}, Matthew J. and {Regan}, John and {Wise}, John H. and {Schive}, Hsi-Yu and {Abel}, Tom and {Emerick}, Andrew and {O'Shea}, Brian W. and {Anninos}, Peter and {Hummels}, Cameron B. and {Khochfar}, Sadegh},
  doi = {10.1093/mnras/stw3291},
  eprint = {1610.09591},
  journal = {\mnras},
  month = {Apr},
  number = {2},
  pages = {2217-2234},
  primaryclass = {astro-ph.CO},
  title = {{GRACKLE: a chemistry and cooling library for astrophysics}},
  volume = {466},
  year = {2017}
}

@article{carbox,
  adsurl = {https://ui.adsabs.harvard.edu/abs/2025arXiv251110558V},
  archiveprefix = {arXiv},
  author = {{Vermari{\"e}n}, Gijs and {Grassi}, Tommaso and {Van de Sande}, Marie and {Viti}, Serena and {Bovino}, Stefano and {Lupi}, Alessandro and {Ruf}, Alexander and {Branca}, Lorenzo and {Walsh}, Catherine},
  doi = {10.48550/arXiv.2511.10558},
  eid = {arXiv:2511.10558},
  eprint = {2511.10558},
  journal = {arXiv e-prints},
  month = {Nov},
  pages = {arXiv:2511.10558},
  primaryclass = {astro-ph.GA},
  title = {{Carbox: an end-to-end differentiable astrochemical simulation framework}},
  year = {2025}
}

@article{branca:2023,
  adsurl = {https://ui.adsabs.harvard.edu/abs/2023MNRAS.518.5718B},
  archiveprefix = {arXiv},
  author = {{Branca}, L. and {Pallottini}, A.},
  doi = {10.1093/mnras/stac3512},
  eprint = {2211.15688},
  journal = {\mnras},
  month = {Jan},
  number = {4},
  pages = {5718-5733},
  primaryclass = {astro-ph.GA},
  title = {{Neural networks: solving the chemistry of the interstellar medium}},
  volume = {518},
  year = {2023}
}

@article{nejad:2005,
  adsurl = {https://ui.adsabs.harvard.edu/abs/2005Ap&SS.299....1N},
  author = {{Nejad}, Lida A.~M.},
  doi = {10.1007/s10509-005-2100-z},
  journal = {\apss},
  month = {Sep},
  number = {1},
  pages = {1-29},
  title = {{A Comparison of Stiff ODE Solvers for Astrochemical Kinetics Problems}},
  volume = {299},
  year = {2005}
}

@article{grassi:2021,
  adsurl = {https://ui.adsabs.harvard.edu/abs/2022A&A...668A.139G},
  archiveprefix = {arXiv},
  author = {{Grassi}, T. and {Nauman}, F. and {Ramsey}, J.~P. and {Bovino}, S. and {Picogna}, G. and {Ercolano}, B.},
  doi = {10.1051/0004-6361/202039956},
  eid = {A139},
  eprint = {2104.09516},
  journal = {\aap},
  month = {Dec},
  pages = {A139},
  primaryclass = {astro-ph.IM},
  title = {{Reducing the complexity of chemical networks via interpretable autoencoders}},
  volume = {668},
  year = {2022}
}

@article{holdship:2021,
  adsurl = {https://ui.adsabs.harvard.edu/abs/2021A&A...653A..76H},
  archiveprefix = {arXiv},
  author = {{Holdship}, J. and {Viti}, S. and {Haworth}, T.~J. and {Ilee}, J.~D.},
  doi = {10.1051/0004-6361/202140357},
  eid = {A76},
  eprint = {2106.14789},
  journal = {\aap},
  month = {Sep},
  pages = {A76},
  primaryclass = {physics.comp-ph},
  title = {{Chemulator: Fast, accurate thermochemistry for dynamical models through emulation}},
  volume = {653},
  year = {2021}
}

@article{sulzer:2023,
  adsurl = {https://ui.adsabs.harvard.edu/abs/2023arXiv231206015S},
  archiveprefix = {arXiv},
  author = {{Sulzer}, Immanuel and {Buck}, Tobias},
  eid = {arXiv:2312.06015},
  eprint = {2312.06015},
  journal = {arXiv e-prints},
  month = {Dec},
  pages = {arXiv:2312.06015},
  primaryclass = {astro-ph.GA},
  title = {{Speeding up astrochemical reaction networks with autoencoders and neural ODEs}},
  year = {2023}
}

@article{asensio24,
  adsurl = {https://ui.adsabs.harvard.edu/abs/2024MNRAS.531.4930A},
  archiveprefix = {arXiv},
  author = {{Asensio Ramos}, A. and {Westendorp Plaza}, C. and {Navarro-Almaida}, D. and {Rivi{\`e}re-Marichalar}, P. and {Wakelam}, V. and {Fuente}, A.},
  doi = {10.1093/mnras/stae1432},
  eprint = {2406.02387},
  journal = {\mnras},
  month = {Jul},
  number = {4},
  pages = {4930-4943},
  primaryclass = {astro-ph.IM},
  title = {{A fast neural emulator for interstellar chemistry}},
  volume = {531},
  year = {2024}
}

@article{gijs25,
  adsurl = {https://ui.adsabs.harvard.edu/abs/2025MLS&T...6b5069V},
  archiveprefix = {arXiv},
  author = {{Vermari{\"e}n}, Gijs and {Bisbas}, Thomas G. and {Viti}, Serena and {Zhao}, Yue and {Tang}, Xuefei and {Ravichandran}, Rahul},
  doi = {10.1088/2632-2153/ade4ee},
  eid = {025069},
  eprint = {2506.14270},
  journal = {Machine Learning: Science and Technology},
  month = {Jun},
  number = {2},
  pages = {025069},
  primaryclass = {astro-ph.GA},
  title = {{NeuralPDR: neural differential equations as surrogate models for photodissociation regions}},
  volume = {6},
  year = {2025}
}

@article{branca24,
  adsurl = {https://ui.adsabs.harvard.edu/abs/2024A&A...684A.203B},
  archiveprefix = {arXiv},
  author = {{Branca}, Lorenzo and {Pallottini}, Andrea},
  doi = {10.1051/0004-6361/202449193},
  eid = {A203},
  eprint = {2402.12435},
  journal = {\aap},
  month = {Apr},
  pages = {A203},
  primaryclass = {astro-ph.GA},
  title = {{Emulating the interstellar medium chemistry with neural operators}},
  volume = {684},
  year = {2024}
}

@article{pelle25,
  adsurl = {https://ui.adsabs.harvard.edu/abs/2025OJAp....8E..96V},
  archiveprefix = {arXiv},
  author = {{van de Bor}, Pelle and {Brennan}, John and {Regan}, John A. and {Mackey}, Jonathan},
  doi = {10.33232/001c.142225},
  eid = {96},
  eprint = {2503.10736},
  journal = {The Open Journal of Astrophysics},
  month = {Jul},
  pages = {96},
  primaryclass = {astro-ph.IM},
  title = {{Bridging Machine Learning and Cosmological Simulations: Using Neural Operators to emulate Chemical Evolution}},
  volume = {8},
  year = {2025}
}

@article{ono25,
  adsurl = {https://ui.adsabs.harvard.edu/abs/2026ApJ...996....9O},
  archiveprefix = {arXiv},
  author = {{Ono}, Sojun and {Sugimura}, Kazuyuki},
  doi = {10.3847/1538-4357/ae1ca9},
  eid = {9},
  eprint = {2508.16114},
  journal = {\apj},
  month = {Jan},
  number = {1},
  pages = {9},
  primaryclass = {astro-ph.GA},
  title = {{Neural-network Chemical Emulator for First-star Formation: Robust Iterative Predictions Over a Wide Density Range}},
  volume = {996},
  year = {2026}
}

@article{codes_2024,
  adsurl = {https://ui.adsabs.harvard.edu/abs/2024arXiv241020886J},
  archiveprefix = {arXiv},
  author = {{Janssen}, Robin and {Sulzer}, Immanuel and {Buck}, Tobias},
  doi = {10.48550/arXiv.2410.20886},
  eid = {arXiv:2410.20886},
  eprint = {2410.20886},
  journal = {arXiv e-prints},
  month = {Oct},
  pages = {arXiv:2410.20886},
  primaryclass = {cs.LG},
  title = {{CODES: Benchmarking Coupled ODE Surrogates}},
  year = {2024}
}

@article{bovino:2016,
  adsurl = {https://ui.adsabs.harvard.edu/abs/2016A&A...590A..15B},
  archiveprefix = {arXiv},
  author = {{Bovino}, S. and {Grassi}, T. and {Capelo}, Pedro R. and {Schleicher}, D.~R.~G. and {Banerjee}, R.},
  doi = {10.1051/0004-6361/201628158},
  eid = {A15},
  eprint = {1510.07016},
  journal = {\aap},
  month = {May},
  pages = {A15},
  primaryclass = {astro-ph.GA},
  title = {{A chemical model for the interstellar medium in galaxies}},
  volume = {590},
  year = {2016}
}

@article{jura:1975,
  adsurl = {https://ui.adsabs.harvard.edu/abs/1975ApJ...197..575J},
  author = {{Jura}, M.},
  doi = {10.1086/153545},
  journal = {\apj},
  month = {May},
  pages = {575-580},
  title = {{Interstellar clouds containing optically thin H$_{2}$.}},
  volume = {197},
  year = {1975}
}

@article{wakelam:2017,
  adsurl = {https://ui.adsabs.harvard.edu/abs/2017MolAs...9....1W},
  archiveprefix = {arXiv},
  author = {{Wakelam}, Valentine and {Bron}, Emeric and {Cazaux}, Stephanie and {Dulieu}, Francois and {Gry}, C{\'e}cile and {Guillard}, Pierre and {Habart}, Emilie and {Hornek{\ae}r}, Liv and {Morisset}, Sabine and {Nyman}, Gunnar and {Pirronello}, Valerio and {Price}, Stephen D. and {Valdivia}, Valeska and {Vidali}, Gianfranco and {Watanabe}, Naoki},
  doi = {10.1016/j.molap.2017.11.001},
  eprint = {1711.10568},
  journal = {Molecular Astrophysics},
  month = {Dec},
  pages = {1-36},
  primaryclass = {astro-ph.GA},
  title = {{H$_{2}$ formation on interstellar dust grains: The viewpoints of theory, experiments, models and observations}},
  volume = {9},
  year = {2017}
}

@article{Glover2010,
  adsurl = {https://ui.adsabs.harvard.edu/abs/2010MNRAS.404....2G},
  archiveprefix = {arXiv},
  author = {{Glover}, S.~C.~O. and {Federrath}, C. and {Mac Low}, M. -M. and {Klessen}, R.~S.},
  doi = {10.1111/j.1365-2966.2009.15718.x},
  eprint = {0907.4081},
  journal = {\mnras},
  month = {May},
  number = {1},
  pages = {2-29},
  primaryclass = {astro-ph.SR},
  title = {{Modelling CO formation in the turbulent interstellar medium}},
  volume = {404},
  year = {2010}
}

@article{Glover2011,
  adsurl = {https://ui.adsabs.harvard.edu/abs/2011MNRAS.412..337G},
  archiveprefix = {arXiv},
  author = {{Glover}, S.~C.~O. and {Mac Low}, M. -M.},
  doi = {10.1111/j.1365-2966.2010.17907.x},
  eprint = {1003.1340},
  journal = {\mnras},
  month = {Mar},
  number = {1},
  pages = {337-350},
  primaryclass = {astro-ph.GA},
  title = {{On the relationship between molecular hydrogen and carbon monoxide abundances in molecular clouds}},
  volume = {412},
  year = {2011}
}

@article{KF1996,
  adsurl = {https://ui.adsabs.harvard.edu/abs/1996ApJS..106..205K},
  author = {{Kingdon}, J.~B. and {Ferland}, G.~J.},
  doi = {10.1086/192335},
  journal = {\apjs},
  month = {Sep},
  pages = {205},
  title = {{Rate Coefficients for Charge Transfer between Hydrogen and the First 30 Elements}},
  volume = {106},
  year = {1996}
}

@article{OConnor2015,
  adsurl = {https://ui.adsabs.harvard.edu/abs/2015ApJS..219....6O},
  archiveprefix = {arXiv},
  author = {{O'Connor}, A.~P. and {Urbain}, X. and {St{\"u}tzel}, J. and {Miller}, K.~A. and {de Ruette}, N. and {Garrido}, M. and {Savin}, D.~W.},
  doi = {10.1088/0067-0049/219/1/6},
  eid = {6},
  eprint = {1408.4696},
  journal = {\apjs},
  month = {Jul},
  number = {1},
  pages = {6},
  primaryclass = {astro-ph.IM},
  title = {{Reaction Studies of Neutral Atomic C with H$_{3}$$^{+}$ using a Merged-beams Apparatus}},
  volume = {219},
  year = {2015}
}

@article{Glover2009,
  adsurl = {https://ui.adsabs.harvard.edu/abs/2009MNRAS.393..911G},
  archiveprefix = {arXiv},
  author = {{Glover}, S.~C.~O. and {Savin}, D.~W.},
  doi = {10.1111/j.1365-2966.2008.14156.x},
  eprint = {0809.0780},
  journal = {\mnras},
  month = {Mar},
  number = {3},
  pages = {911-948},
  primaryclass = {astro-ph},
  title = {{Is H$^{+}$$_{3}$ cooling ever important in primordial gas?}},
  volume = {393},
  year = {2009}
}

@article{wakelam:2012,
  adsurl = {https://ui.adsabs.harvard.edu/abs/2012ApJS..199...21W},
  archiveprefix = {arXiv},
  author = {{Wakelam}, V. and {Herbst}, E. and {Loison}, J. -C. and {Smith}, I.~W.~M. and {Chandrasekaran}, V. and {Pavone}, B. and {Adams}, N.~G. and {Bacchus-Montabonel}, M. -C. and {Bergeat}, A. and {B{\'e}roff}, K. and {Bierbaum}, V.~M. and {Chabot}, M. and {Dalgarno}, A. and {van Dishoeck}, E.~F. and {Faure}, A. and {Geppert}, W.~D. and {Gerlich}, D. and {Galli}, D. and {H{\'e}brard}, E. and {Hersant}, F. and {Hickson}, K.~M. and {Honvault}, P. and {Klippenstein}, S.~J. and {Le Picard}, S. and {Nyman}, G. and {Pernot}, P. and {Schlemmer}, S. and {Selsis}, F. and {Sims}, I.~R. and {Talbi}, D. and {Tennyson}, J. and {Troe}, J. and {Wester}, R. and {Wiesenfeld}, L.},
  doi = {10.1088/0067-0049/199/1/21},
  eid = {21},
  eprint = {1201.5887},
  journal = {\apjs},
  month = {Mar},
  number = {1},
  pages = {21},
  primaryclass = {astro-ph.GA},
  title = {{A KInetic Database for Astrochemistry (KIDA)}},
  volume = {199},
  year = {2012}
}

@article{OSU1980,
  adsurl = {https://ui.adsabs.harvard.edu/abs/1980ApJS...43....1P},
  author = {{Prasad}, S.~S. and {Huntress}, Jr., W.~T.},
  doi = {10.1086/190665},
  journal = {\apjs},
  month = {May},
  pages = {1-35},
  title = {{A model for gas phase chemistry in interstellar clouds: I. The basic model, library of chemical reactions, and chemistry among C, N, and O compounds.}},
  volume = {43},
  year = {1980}
}

@article{UMIST1991,
  adsurl = {https://ui.adsabs.harvard.edu/abs/1991A&A...242..241M},
  author = {{Millar}, T.~J.},
  journal = {\aap},
  month = {Feb},
  pages = {241},
  title = {{Phosphorus chemistry in dense interstellar clouds.}},
  volume = {242},
  year = {1991}
}

@article{shen:2013,
  adsurl = {https://ui.adsabs.harvard.edu/abs/2013ApJ...765...89S},
  archiveprefix = {arXiv},
  author = {{Shen}, Sijing and {Madau}, Piero and {Guedes}, Javiera and {Mayer}, Lucio and {Prochaska}, J. Xavier and {Wadsley}, James},
  doi = {10.1088/0004-637X/765/2/89},
  eid = {89},
  eprint = {1205.0270},
  journal = {\apj},
  month = {Mar},
  number = {2},
  pages = {89},
  primaryclass = {astro-ph.CO},
  title = {{The Circumgalactic Medium of Massive Galaxies at z \raisebox{-0.5ex}\textasciitilde 3: A Test for Stellar Feedback, Galactic Outflows, and Cold Streams}},
  volume = {765},
  year = {2013}
}

@article{Sobol1967,
  author = {I. M. Sobol},
  doi = {10.1016/0041-5553(67)90144-9},
  journal = {U.S.S.R. Comput. Math. Math. Phys.},
  note = {Translated from Zh. Vychisl. Mat. Mat. Fiz., 7(4):784--802, 1967},
  number = {4},
  pages = {86--112},
  title = {On the distribution of points in a cube and the approximate evaluation of integrals},
  volume = {7},
  year = {1967}
}

@article{lu_learning_2021,
  adsurl = {https://www.nature.com/articles/s42256-021-00302-5},
  author = {Lu, Lu and Jin, Pengzhan and Pang, Guofei and Zhang, Zhongqiang and Karniadakis, George Em},
  doi = {10.1038/s42256-021-00302-5},
  issn = {2522-5839},
  journal = {Nature Machine Intelligence},
  number = {3},
  pages = {218--229},
  publisher = {Nature Publishing Group},
  rights = {2021 The Author(s), under exclusive licence to Springer Nature Limited},
  shortjournal = {Nat Mach Intell},
  title = {Learning nonlinear operators via {DeepONet} based on the universal approximation theorem of operators},
  volume = {3},
  year = {2021}
}

@article{lu_2022_comprehensive,
  adsurl = {https://ui.adsabs.harvard.edu/abs/2022CMAME.39314778L},
  archiveprefix = {arXiv},
  author = {{Lu}, Lu and {Meng}, Xuhui and {Cai}, Shengze and {Mao}, Zhiping and {Goswami}, Somdatta and {Zhang}, Zhongqiang and {Karniadakis}, George Em},
  doi = {10.1016/j.cma.2022.114778},
  eid = {114778},
  eprint = {2111.05512},
  journal = {Computer Methods in Applied Mechanics and Engineering},
  month = {Apr},
  pages = {114778},
  primaryclass = {physics.comp-ph},
  title = {{A comprehensive and fair comparison of two neural operators (with practical extensions) based on FAIR data}},
  volume = {393},
  year = {2022}
}

@inproceedings{chen_neural_2018,
  adsurl = {https://proceedings.neurips.cc/paper_files/paper/2018/hash/69386f6bb1dfed68692a24c8686939b9-Abstract.html},
  author = {Chen, Ricky T. Q. and Rubanova, Yulia and Bettencourt, Jesse and Duvenaud, David},
  booktitle = {Proceedings of the 32nd International Conference on Neural Information Processing Systems},
  date = {2018-12-03},
  location = {Red Hook, {NY}, {USA}},
  pages = {6572--6583},
  title = {Neural ordinary differential equations},
  urldate = {2026-02-19},
  year = {2018}
}

@inproceedings{paszke_pytorch_2019,
  adsurl = {https://proceedings.neurips.cc/paper_files/paper/2019/hash/bdbca288fee7f92f2bfa9f7012727740-Abstract.html},
  author = {Paszke, Adam and Gross, Sam and Massa, Francisco and Lerer, Adam and Bradbury, James and Chanan, Gregory and Killeen, Trevor and Lin, Zeming and Gimelshein, Natalia and Antiga, Luca and Desmaison, Alban and Köpf, Andreas and Yang, Edward and {DeVito}, Zach and Raison, Martin and Tejani, Alykhan and Chilamkurthy, Sasank and Steiner, Benoit and Fang, Lu and Bai, Junjie and Chintala, Soumith},
  booktitle = {Proceedings of the 33rd International Conference on Neural Information Processing Systems},
  location = {Red Hook, {NY}, {USA}},
  pages = {8026--8037},
  title = {{PyTorch}: an imperative style, high-performance deep learning library},
  year = {2019}
}

@inproceedings{akiba2019optuna,
  adsurl = {https://dl.acm.org/doi/10.1145/3292500.3330701},
  author = {Akiba, Takuya and Sano, Shotaro and Yanase, Toshihiko and Ohta, Takeru and Koyama, Masanori},
  booktitle = {Proceedings of the 25th {ACM} {SIGKDD} International Conference on Knowledge Discovery \& Data Mining},
  date = {2019-07-25},
  doi = {10.1145/3292500.3330701},
  isbn = {978-1-4503-6201-6},
  location = {New York, {NY}, {USA}},
  pages = {2623--2631},
  series = {{KDD} '19},
  shorttitle = {Optuna},
  title = {Optuna: A Next-generation Hyperparameter Optimization Framework},
  year = {2019}
}

@inproceedings{defazio_road_2024,
  adsurl = {https://proceedings.neurips.cc/paper_files/paper/2024/hash/136b9a13861308c8948cd308ccd02658-Abstract-Conference.html},
  author = {Defazio, Aaron and Yang, Xingyu (Alice) and Mehta, Harsh and Mishchenko, Konstantin and Khaled, Ahmed and Cutkosky, Ashok},
  booktitle = {Proceedings of the 38th International Conference on Neural Information Processing Systems},
  date = {2024-12-10},
  isbn = {979-8-3313-1438-5},
  location = {Red Hook, {NY}, {USA}},
  pages = {9974--10007},
  title = {The road less scheduled},
  volume = {37},
  year = {2024}
}

@article{lienen2022torchode,
  adsurl = {https://ui.adsabs.harvard.edu/abs/2022arXiv221012375L},
  archiveprefix = {arXiv},
  author = {{Lienen}, Marten and {G{\"u}nnemann}, Stephan},
  doi = {10.48550/arXiv.2210.12375},
  eid = {arXiv:2210.12375},
  eprint = {2210.12375},
  journal = {arXiv e-prints},
  month = {Oct},
  pages = {arXiv:2210.12375},
  primaryclass = {cs.LG},
  title = {{torchode: A Parallel ODE Solver for PyTorch}},
  year = {2022}
}

@article{deb_fast_2002,
  adsurl = {https://ui.adsabs.harvard.edu/abs/2002ITEC....6..182D},
  author = {{Deb}, Kalyanmoy and {Pratap}, Amrit and {Agarwal}, Sameer and {Meyarivan}, T.},
  doi = {10.1109/4235.996017},
  journal = {IEEE Transactions on Evolutionary Computation},
  month = {Jan},
  number = {2},
  pages = {182},
  title = {{A Fast and Elitist Multiobjective Genetic Algorithm: NSGA-II}},
  volume = {6},
  year = {2002}
}

@inproceedings{lakshminarayanan_simple_2017,
  adsurl = {https://dl.acm.org/doi/10.5555/3295222.3295387},
  author = {Lakshminarayanan, Balaji and Pritzel, Alexander and Blundell, Charles},
  booktitle = {Proceedings of the 31st International Conference on Neural Information Processing Systems},
  isbn = {978-1-5108-6096-4},
  location = {Red Hook, {NY}, {USA}},
  pages = {6405--6416},
  title = {Simple and scalable predictive uncertainty estimation using deep ensembles},
  year = {2017}
}

@article{fronk_vanishing_2025,
  adsurl = {https://ui.adsabs.harvard.edu/abs/2025Chaos..35k3113F},
  archiveprefix = {arXiv},
  author = {{Fronk}, Colby and {Petzold}, Linda},
  doi = {10.1063/5.0294317},
  eid = {113113},
  eprint = {2508.01519},
  journal = {Chaos},
  month = {Nov},
  number = {11},
  pages = {113113},
  primaryclass = {stat.ML},
  title = {{The vanishing gradient problem for stiff neural differential equations}},
  volume = {35},
  year = {2025}
}

@article{silke24,
  adsurl = {https://ui.adsabs.harvard.edu/abs/2024ApJ...969...79M},
  archiveprefix = {arXiv},
  author = {{Maes}, Silke and {De Ceuster}, Frederik and {Van de Sande}, Marie and {Decin}, Leen},
  doi = {10.3847/1538-4357/ad47a1},
  eid = {79},
  eprint = {2405.03274},
  journal = {\apj},
  month = {Jul},
  number = {2},
  pages = {79},
  primaryclass = {physics.comp-ph},
  title = {{MACE: A Machine-learning Approach to Chemistry Emulation}},
  volume = {969},
  year = {2024}
}

@inproceedings{bergstra_algorithms_2011,
  adsurl = {https://papers.nips.cc/paper_files/paper/2011/hash/86e8f7ab32cfd12577bc2619bc635690-Abstract.html},
  author = {Bergstra, James and Bardenet, Rémi and Bengio, Yoshua and Kégl, Balázs},
  booktitle = {Proceedings of the 25th International Conference on Neural Information Processing Systems},
  date = {2011-12-12},
  isbn = {978-1-61839-599-3},
  location = {Red Hook, {NY}, {USA}},
  pages = {2546--2554},
  title = {Algorithms for hyper-parameter optimization},
  year = {2011}
}

@article{li_hyperband_2017,
  adsurl = {http://jmlr.org/papers/v18/16-558.html},
  author = {Li, Lisha and Jamieson, Kevin and {DeSalvo}, Giulia and Rostamizadeh, Afshin and Talwalkar, Ameet},
  issn = {1533-7928},
  journal = {Journal of Machine Learning Research},
  number = {185},
  pages = {1--52},
  title = {Hyperband: A Novel Bandit-Based Approach to Hyperparameter Optimization},
  volume = {18},
  year = {2018}
}

@misc{odepack,
       author = {{Hindmarsh}, Alan C.},
        title = "{ODEPACK: Ordinary differential equation solver library}",
 howpublished = {Astrophysics Source Code Library},
         year = 2019,
        month = may,
          eid = {ascl:1905.021},
archivePrefix = {ascl},
       eprint = {1905.021},
       adsurl = {https://ui.adsabs.harvard.edu/abs/2019ascl.soft05021H},
      note = {record \href{https://ui.adsabs.harvard.edu/abs/2019ascl.soft05021H}{ascl:1905.021}}
      
}

@misc{jax2018github,
       author = {{Bradbury}, James and {Frostig}, Roy and {Hawkins}, Peter and {Johnson}, Matthew James and {Leary}, Chris and {Maclaurin}, Dougal and {Necula}, George and {Paszke}, Adam and {VanderPlas}, Jake and {Wanderman-Milne}, Skye and et al.},
        title = "{JAX: Autograd and XLA}",
 howpublished = {Astrophysics Source Code Library},
         year = 2021,
        month = nov,
          eid = {ascl:2111.002},
archivePrefix = {ascl},
       eprint = {2111.002},
       adsurl = {https://ui.adsabs.harvard.edu/abs/2021ascl.soft11002B},
      note = {record \href{https://ui.adsabs.harvard.edu/abs/2021ascl.soft11002B}{ascl:2111.002}}
}

@misc{storcks_astronomix_2025,
  doi = {10.5281/ZENODO.17782162},
  url = {https://zenodo.org/doi/10.5281/zenodo.17782162},
  author = {Storcks, Leonard},
  title = {astronomix - differentiable MHD in JAX},
  year = {2025},
  copyright = {MIT License},
  note = {Zenodo, DOI: \href{https://zenodo.org/doi/10.5281/zenodo.17782162}{10.5281/zenodo.17782162}}
}

\begin{appendix}

\section{Hyperparameter optimization details and results}
\label{app:hyperparams}

This appendix provides supplementary details on the HPO procedure used in CODES, including the optimization setup, the selection of Pareto-optimal configurations, and additional results and tables supporting Sects. \ref{subsec:tuning} and  \ref{subsec:hyperparam_results}.

\subsection{Optimization setup and algorithms}
\label{app:hyperparams:setup}

Given the high computational cost of the tuning runs, CODES employs the \textsc{Optuna} framework \citep{akiba2019optuna} to efficiently explore heterogeneous hyperparameter search spaces. In particular, CODES leverages \textsc{Optuna}'s adaptive sampling strategies, pruning of unpromising trials, and asynchronous parallel execution to reduce computational overhead while maintaining robust optimization performance.

For adaptive sampling, CODES uses Optuna samplers that propose new configurations based on outcomes of previous trials. For single-objective optimization, CODES employs the \texttt{TPESampler}, based on the Tree-Parzen estimator approach \citep{bergstra_algorithms_2011}. For the dual-objective optimization used throughout this work, CODES employs \texttt{NSGAIISampler}, implementing the NSGA-II genetic algorithm \citep{deb_fast_2002}. We used a fixed population size of 50 for NSGA-II, while all other sampler parameters (including crossover and mutation settings) are left at \textsc{Optuna}’s default values. All reported Pareto fronts and selected configurations in this paper are obtained using \texttt{NSGAIISampler}.

To reduce wasted compute on poor configurations, CODES uses \textsc{Optuna} pruners to terminate unpromising trials early. By default, CODES employs the \texttt{HyperbandPruner} \citep{li_hyperband_2017}, which dynamically allocates computational resources across trials and can substantially reduce overall cost compared to running all trials to completion. However, \textsc{Optuna}’s built-in pruning mechanisms are only directly compatible with single-objective optimization. To nevertheless retain some of the benefits of early stopping in our dual-objective setup, we implemented a lightweight heuristic: during tuning, we tracked the running mean and standard deviation of trial durations and pruned trials exceeding the mean duration by more than one standard deviation after an initial warmup phase.
This strategy is motivated by the observation that, within a given architecture and dataset, unusually long training times typically arise from configurations with increased model complexity or solver overhead, which also tend to incur higher inference cost. 
Pruning only the slowest trials therefore yields a disproportionate reduction in computational cost, as these configurations dominate the overall resource consumption by definition. 
We emphasize that this heuristic is not guaranteed to preserve all Pareto-optimal configurations, but serves as a lightweight mechanism to eliminate clearly inefficient trials in the absence of native multi-objective pruning support.

Finally, \textsc{Optuna} supports asynchronous parallel execution of trials, allowing multiple configurations to be evaluated concurrently. While this does not reduce total compute requirements, it substantially reduces wall-clock time.

The tuning budget for each study was set heuristically to $15 \times N_h$, where $N_h$ denotes the number of tuned hyperparameters. Across studies, $N_h$ ranges from 11 to 17 depending on the architecture and on whether the dataset includes additional physical parameters (see Sect. \ref{sec:surrogates}), resulting in 165 to 255 trials per study. To control cost, we used a trial length of 2000 epochs during the main tuning stage. After selecting Pareto-optimal configurations, we additionally performed a fine-tuning stage with 8000 epochs, restricting the search to hyperparameters that directly couple to training duration (learning rate, weight decay, and, where applicable, schedule or optimizer coefficients). This fine-tuning stage accounts for the fact that the computational budget of individual trials is typically lower than for the final training run, which can bias the optimal values of hyperparameters whose effect depends on training duration. The number of fine-tuning trials was set to $10 \times N_h$ (typically 30--40 trials per study).

\begin{figure}[t]
  \centering

  \includegraphics[width=0.5\textwidth]{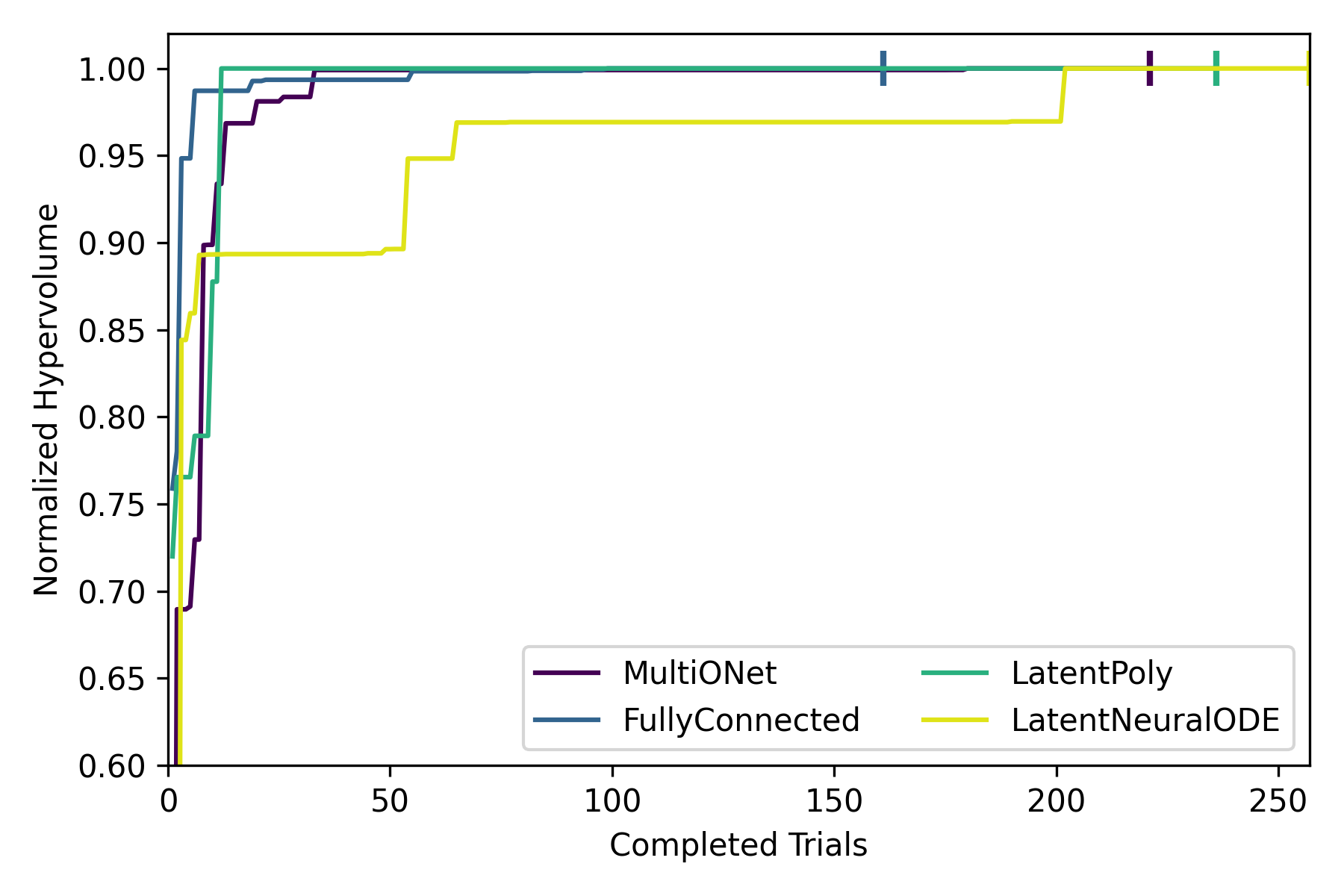}
  \par\medskip
  \includegraphics[width=0.5\textwidth]{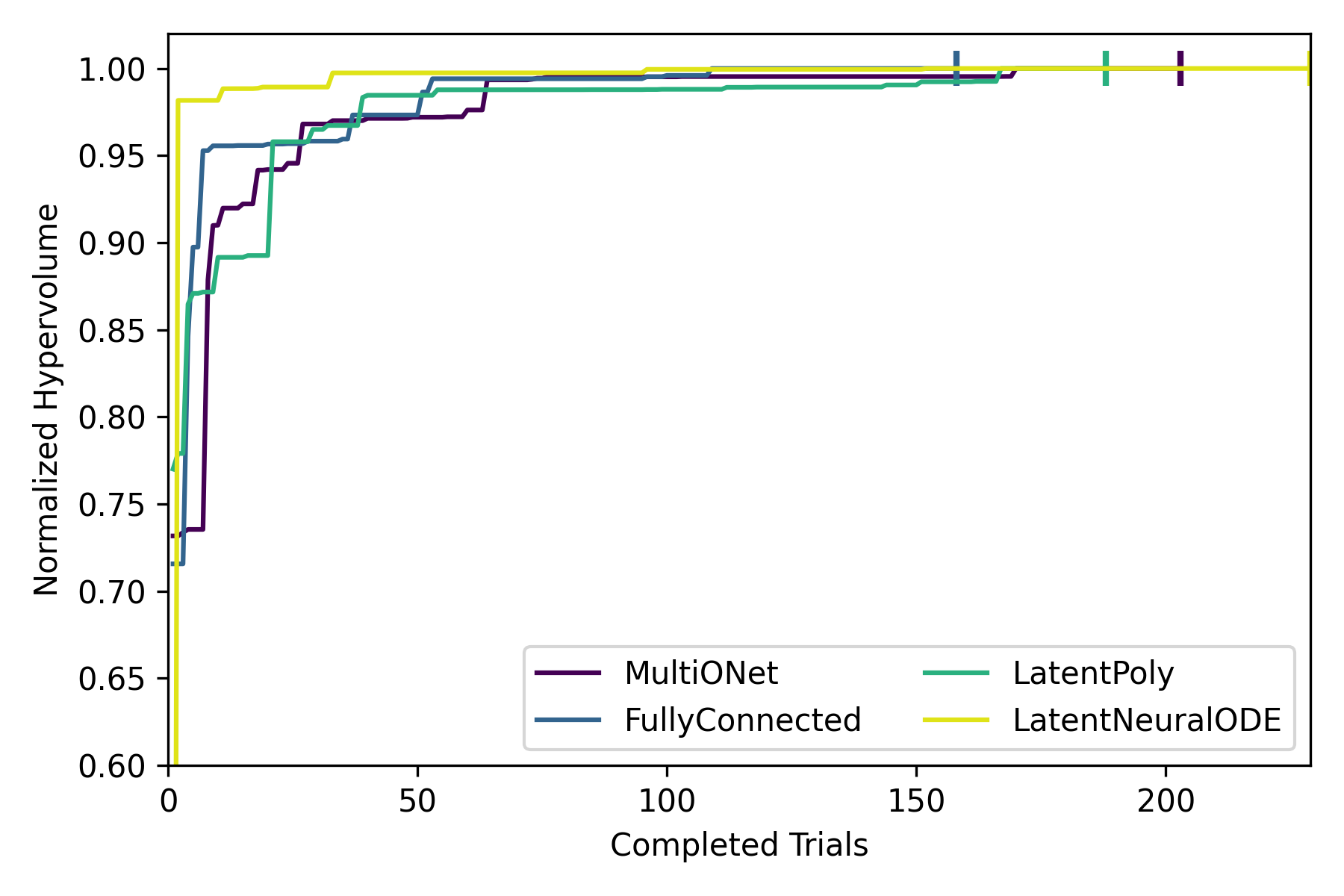}
  \par\medskip
  \includegraphics[width=0.5\textwidth]{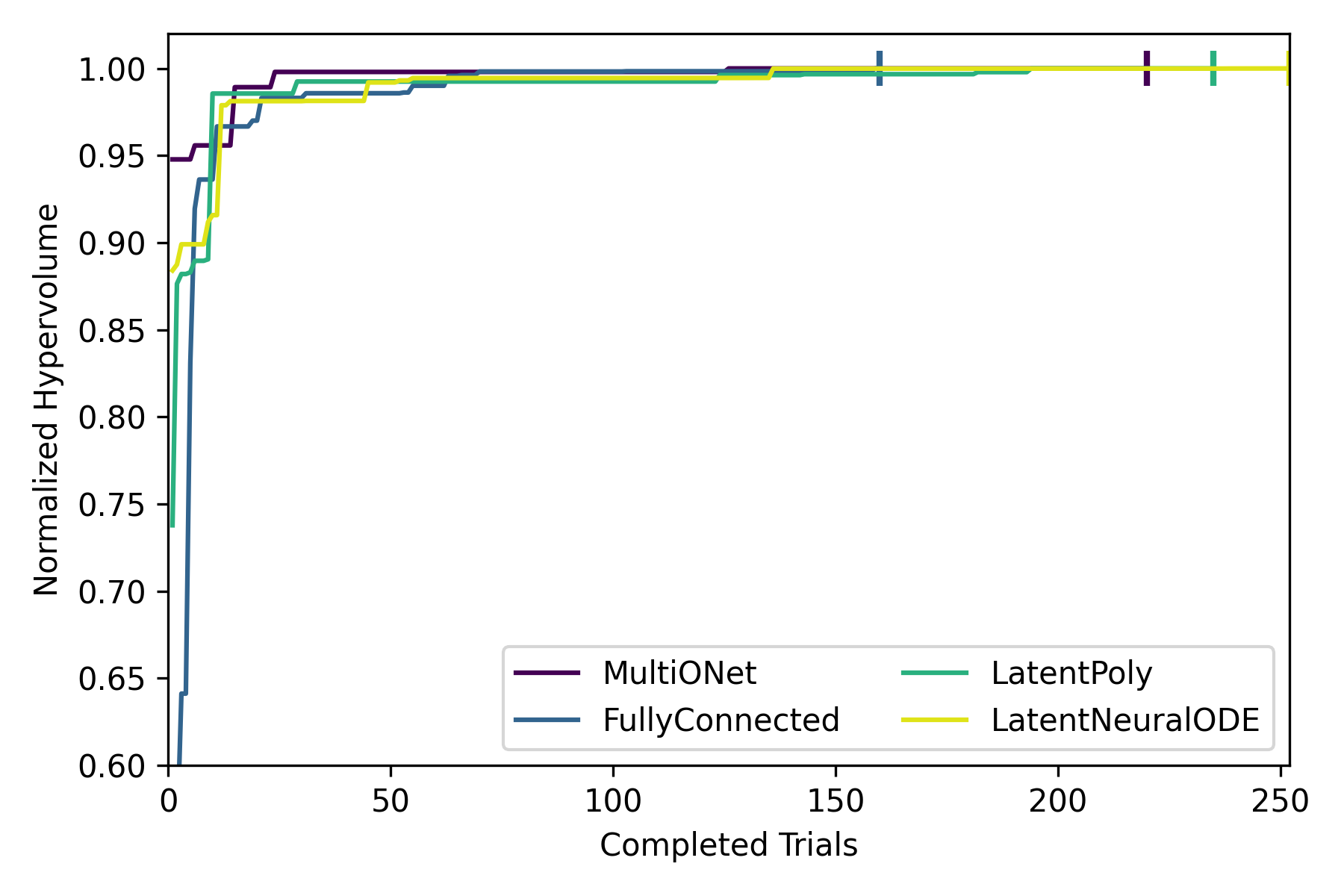}

  \caption{Normalized hypervolume evolution during HPO for the
  primordial parametric (top), cloud (middle), and cloud parametric (bottom) datasets.}
  \label{fig:hypervolume_appendix}
\end{figure}

\subsection{Selection of Pareto-optimal configurations}
\label{app:hyperparams:selection}

For each dataset--architecture pair, dual-objective optimization yields a set of Pareto-optimal trials in the two-dimensional objective space spanned by LAE$_{99}$ and inference time (see Sect. \ref{subsec:tuning}). From each Pareto front, we selected one configuration for downstream training and evaluation. The selection was performed manually by choosing a configuration near the knee of the Pareto curve, prioritizing substantial efficiency gains while avoiding disproportionate losses in accuracy.

In principle, this step could be automated via knee-finding heuristics or scalarization schemes. In practice, such automation offers limited benefit in our setting for two reasons. First, the curvature and scaling of Pareto fronts differ between architectures and datasets, making a single, scale-invariant selection rule difficult to define. Second, Pareto fronts are obtained from stochastic optimization with finite budgets. Automated knee-finding can be sensitive to sampling noise and small perturbations of the front. We therefore adopted manual knee selection as a robust and transparent procedure that aligns with the practical surrogate-selection question of interest: selecting one of the most accurate configurations that does not sacrifice efficiency disproportionately.

\begin{table}[h]
\centering
\caption{Accuracy–efficiency trade-off between the lowest-error and Pareto-selected configurations (see Fig. \ref{fig:pareto_fronts}).}
\label{tab:hpo_tradeoff}
\small
\begin{tabular}{l l r r r}
\toprule
Dataset & Arch & $\Delta$Error $\downarrow$ [\%] & $\Delta$Time $\uparrow$ [\%] & Ratio R \\
\midrule
Primordial   & MON   & 24.9 & 155 & 6.2 \\
             & FCNN  & 24.0 & 417 & 17.4 \\
             & LP    & 16.7 & 148 & 8.9 \\
             & LNODE & 13.9 & 98  & 7.1 \\
\midrule
Primordial   & MON   & 29.6 & 109 & 3.7 \\
Parametric   & FCNN  & 37.3 & 265 & 7.1 \\
\midrule
Cloud        & LP    & 9.5  & 143 & 15.1 \\
             & LNODE & 1.7  & 70  & 41.2 \\
\midrule
Cloud        & MON   & 4.6  & 36  & 7.8 \\
Parametric   & LP    & 5.4  & 411 & 76.1 \\
             & LNODE & 1.6  & 28  & 17.5 \\
\midrule
\multicolumn{2}{l}{\textbf{Averages}} & \textbf{15.4} & \textbf{171} & \textbf{18.9} \\ 
\bottomrule
\end{tabular}
\tablefoot{Architectures where both configurations coincide are omitted. 
Reported values indicate relative percentage changes of the accuracy-only configuration with respect to the Pareto-selected one. 
The final column reports the ratio $R = \Delta \text{Time} / \Delta \text{Error}$, i.e., the percentage increase in inference time required to achieve a 1\% reduction in error.}
\end{table}

\subsection{Pareto-front area evolution across datasets}
\label{app:hyperparams:hypervolume}

To provide a post-hoc indication of optimization progress within the allocated budgets, we monitor the evolution of the Pareto-front area (hypervolume) over the course of each study, computed with respect to a reference point slightly worse than the worst observed configuration in both objectives (see Sect. \ref{subsec:hyperparam_results}). Figure \ref{fig:hypervolume_appendix} shows the normalized area evolution for the datasets primordial parametric, cloud and cloud parametric datasets, respectively.

\begin{center}
\begin{minipage}{\columnwidth}
\centering

\captionof{table}{Chosen hyperparameters for MultiONet.}
\label{tab:hyperparams:a}
\scriptsize
\begin{tabular}{p{1.6cm} C{1.3cm} C{1.3cm} C{1.3cm} C{1.3cm}}
\toprule
Hyperparameter & Cloud & Cloud Parametric & Primordial & Primordial Parametric \\
\midrule
Scheduler & poly & poly & schedulefree & poly \\
Optimizer & AdamW & AdamW & AdamW & AdamW \\
Loss & MSE & MSE & SmoothL1 & MSE \\
Activation & Mish & ELU & LeakyReLU & PReLU \\
Hidden layers B & 8 & 4 & 3 & 5 \\
Hidden size & 100 & 100 & 160 & 50 \\
Output factor & 50 & 40 & 130 & 74 \\
Hidden layers T & 1 & 4 & 2 & 1 \\
Params branch & -- & True & -- & True \\
Poly power & 0.647 & 8.20 & -- & 2.57 \\
Learning rate & 3.7e-03 & 3.2e-03 & 2.7e-03 & 5.2e-03 \\
Reg. factor & 4.6e-05 & 6.5e-03 & 0.117 & 1.4e-03 \\
Beta & -- & -- & 0.732 & -- \\
\bottomrule
\end{tabular}

\vspace{0.35cm}

\captionof{table}{Chosen hyperparameters for LatentNeuralODE.}
\label{tab:hyperparams:b}
\scriptsize
\begin{tabular}{p{1.6cm} C{1.3cm} C{1.3cm} C{1.3cm} C{1.3cm}}
\toprule
Hyperparameter & Cloud & Cloud Parametric & Primordial & Primordial Parametric \\
\midrule
Scheduler & poly & schedulefree & schedulefree & cosine \\
Optimizer & SGD & SGD & SGD & SGD \\
Loss & MSE & SmoothL1 & MSE & MSE \\
Activation & SiLU & GELU & SiLU & SiLU \\
Coder layers & 8 & 2 & 1 & 1 \\
Coder width & 80 & 180 & 580 & 360 \\
Latent features & 1 & 10 & 9 & 10 \\
ODE layers & 3 & 3 & 6 & 8 \\
ODE width & 250 & 220 & 480 & 220 \\
ODE tanh reg & False & False & True & True \\
Encode params & -- & False & -- & False \\
Poly power & 3.43 & -- & -- & -- \\
Momentum & 0.143 & 0.075 & 0.274 & 0.331 \\
Learning rate & 3.9e-05 & 4.5e-03 & 5.9e-04 & 2.8e-06 \\
Reg. factor & 3.41 & 3.6e-04 & 9.2e-03 & 6.4e-05 \\
Eta min & -- & -- & -- & 5.3e-04 \\
Beta & -- & 14.4 & -- & -- \\
\bottomrule
\end{tabular}

\vspace{0.35cm}

\captionof{table}{Chosen hyperparameters for FullyConnected.}
\label{tab:hyperparams:c}
\scriptsize
\begin{tabular}{p{1.6cm} C{1.3cm} C{1.3cm} C{1.3cm} C{1.3cm}}
\toprule
Hyperparameter & Cloud & Cloud Parametric & Primordial & Primordial Parametric \\
\midrule
Scheduler & cosine & poly & poly & poly \\
Optimizer & AdamW & AdamW & AdamW & AdamW \\
Loss & SmoothL1 & SmoothL1 & SmoothL1 & SmoothL1 \\
Activation & PReLU & ELU & ELU & Mish \\
Hidden size & 310 & 290 & 470 & 470 \\
Hidden layers & 1 & 5 & 5 & 3 \\
Poly power & -- & 8.16 & 3.59 & 0.163 \\
Eta min & 2.7e-03 & -- & -- & -- \\
Beta & 8.39 & 2.66 & 0.863 & 39.5 \\
Learning rate & 2.2e-03 & 0.0122 & 1.8e-03 & 1.3e-04 \\
Reg. factor & 3.4e-04 & 0.114 & 5.3e-05 & 0.0333 \\
\bottomrule
\end{tabular}

\vspace{0.35cm}

\captionof{table}{Chosen hyperparameters for LatentPoly.}
\label{tab:hyperparams:d}
\scriptsize
\begin{tabular}{p{1.6cm} C{1.3cm} C{1.3cm} C{1.3cm} C{1.3cm}}
\toprule
Hyperparameter & Cloud & Cloud Parametric & Primordial & Primordial Parametric \\
\midrule
Scheduler & poly & poly & schedulefree & poly \\
Optimizer & AdamW & AdamW & SGD & AdamW \\
Loss & MSE & SmoothL1 & MSE & MSE \\
Activation & GELU & ReLU & ReLU & ELU \\
Coder layers & 2 & 1 & 2 & 1 \\
Coder width & 290 & 170 & 470 & 700 \\
Latent features & 2 & 10 & 8 & 7 \\
Degree & 6 & 9 & 4 & 3 \\
Coeff network & -- & False & -- & False \\
Poly power & 5.18 & 2.46 & -- & 0.264 \\
Momentum & -- & -- & 0.0161 & -- \\
Learning rate & 9.3e-05 & 8.2e-03 & 2.4e-04 & 6.0e-06 \\
Reg. factor & 2.1e-04 & 0.0567 & 1.1e-03 & 0.0510 \\
\bottomrule
\end{tabular}

\end{minipage}
\end{center}

\subsection{Accuracy--efficiency trade-off statistics}
\label{app:hyperparams:tradeoff}

Table \ref{tab:hpo_tradeoff} quantifies the accuracy--efficiency trade-off between the Pareto-selected and accuracy-only (lowest-error) configurations for all studies where these differ. It reports relative percentage changes in LAE$_{99}$ and inference time, as well as the ratio
$R = \Delta \text{Time} / \Delta \text{Error}$, interpreted as the percentage increase in inference time required to achieve a 1\% reduction in error.

\subsection{Optimized hyperparameters}
\label{app:hyperparams:tables}

Tables \ref{tab:hyperparams:a}, \ref{tab:hyperparams:b}, \ref{tab:hyperparams:c}, and \ref{tab:hyperparams:d} report the hyperparameter values selected from the Pareto fronts for each architecture and dataset. Hyperparameter names follow the notation used in the CODES configuration files. Further implementation details are provided in the CODES documentation\footnote{\href{https://astroai-lab.de/CODES-Benchmark/}{https://astroai-lab.de/CODES-Benchmark/}.}. The optimized hyperparameters fall into two categories: shared training hyperparameters, which are common across architectures and architecture-specific hyperparameters, which control structural choices such as layer widths, depths, latent dimensionality, or parameter handling. Entries marked as “–” indicate that a given hyperparameter is not applicable to the corresponding architecture.

\section{Additional accuracy diagnostics}
\label{app:additional_accuracy}

The additional diagnostics in Table \ref{table:results_additional} provide context for the tail behavior of surrogate errors and for temperature-specific accuracy. 
Across all datasets and architectures, the maximum log-space absolute error (LAE$_{100}$) is substantially larger than LAE$_{99}$, typically exceeding it by factors of two to more than an order of magnitude. 
These values correspond to extreme outliers, with errors spanning several to tens of orders of magnitude, and confirm that worst-case failures are indeed catastrophic in absolute terms. 
While models with lower LAE$_{99}$ generally also exhibit lower LAE$_{100}$, this relationship is loose and non-monotonic, underscoring the statistical instability of the maximum error as a performance metric. 
The magnitude of LAE$_{100}$ appears to be primarily determined by the dataset rather than the architecture, with more complex datasets exhibiting higher worst-case errors, consistent with their larger dynamic ranges.

In contrast, the temperature-only error (Temp. LAE) is consistently small across all architectures and datasets, and typically far below the corresponding mLAE values. 
This reflects the comparatively limited dynamic range of temperature relative to chemical abundances and indicates that large chemical outliers do not necessarily translate into large temperature errors. 
From an application perspective, this is encouraging, as temperature evolution is often the primary quantity of interest in coupled hydrodynamical--chemical simulations.

\begin{table}[t]
\centering
\caption{Additional accuracy diagnostics.}
\label{table:results_additional}
\small
\setlength{\tabcolsep}{0pt}
\renewcommand{\arraystretch}{1.0}
\begin{tabular}{p{1.6cm} p{1.4cm} 
                C{1.5cm} C{1.4cm} C{1.4cm} C{1.6cm}}
\toprule
Dataset & Arch & mLAE & LAE$_{99}$ & LAE$_{100}$ & Temp. LAE \\
 & & [dex] & [dex] & [dex] & [dex] \\
\midrule
Primordial & MON   & 0.343 & 2.47 & 6.24 & 0.126 \\
           & FCNN  & \textbf{0.094} & \textbf{0.693} & \textbf{3.47} & \textbf{0.030} \\
           & LNODE & 0.661 & 3.04 & 6.88 & 0.203 \\
           & LP    & 0.843 & 3.35 & 7.07 & 0.27 \\
\midrule
Primordial & MON   & 0.613 & 5.96 & 18.5 & 0.126 \\
Parametric & FCNN  & \textbf{0.259} & \textbf{1.74} & 14.5 & \textbf{0.060} \\
           & LNODE & 0.693 & 3.32 & 14.8 & 0.149 \\
           & LP    & 0.743 & 3.43 & \textbf{14.2} & 0.17 \\
\midrule
Cloud & MON   & \textbf{0.034} & \textbf{0.326} & 6.36 & 0.008 \\
      & FCNN  & 0.035 & 0.384 & \textbf{4.33} & \textbf{0.005} \\
      & LNODE & 1.92 & 4.55 & 8.64 & 0.993 \\
      & LP    & 1.64 & 4.70 & 8.36 & 0.497 \\
\midrule
Cloud        & MON   & \textbf{0.121} & \textbf{0.831} & \textbf{9.43} & \textbf{0.017} \\
Parametric   & FCNN  & 0.244 & 2.80 & 16.9 & 0.019 \\
             & LNODE & 1.51 & 4.80 & 14.7 & 0.253 \\
             & LP    & 1.37 & 4.51 & 13.5 & 0.274 \\
\bottomrule
\end{tabular}
\tablefoot{In addition to mLAE and LAE$_{99}$, we report the maximum log-space absolute error (LAE$_{100}$) and the  temperature-only absolute error (Temp.\ LAE). Best values per dataset are shown in bold.}
\end{table}

\section{Catastrophic error detection}
\label{app:error_detection}

\begin{figure*}
  \centering
  \includegraphics[width=0.9\textwidth]{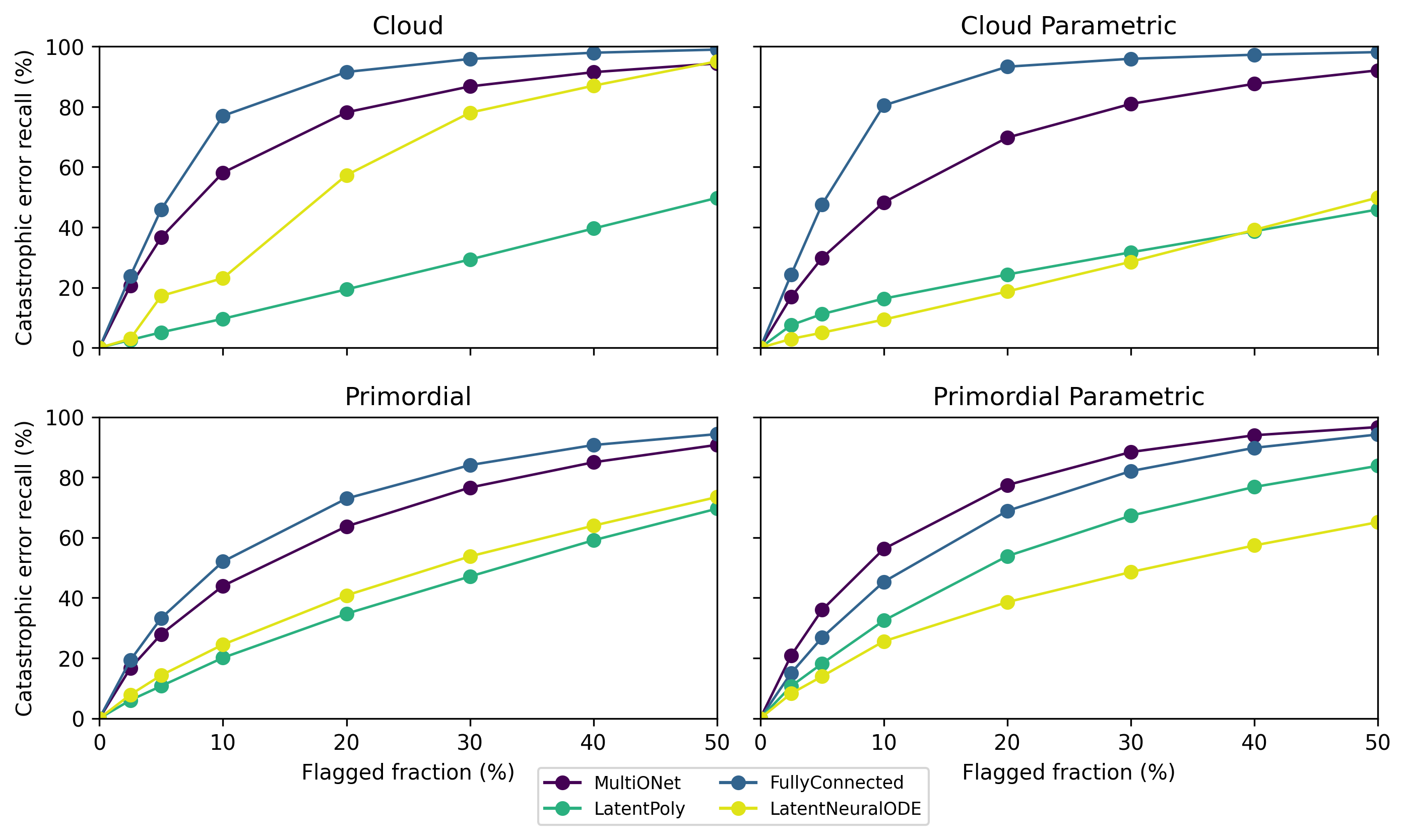}
\caption{Same as Fig. \ref{fig:uq_detection} but defining catastrophic errors as those exceeding the 90th percentile log-space error, illustrating the robustness of the detection trends to the chosen threshold.}
  \label{fig:uq_detection_90}
\end{figure*}

In Sect. \ref{subsec:UQ}, catastrophic prediction errors are defined as those exceeding the LAE$_{99}$, and detection efficiency is evaluated relative to this threshold.
To assess whether the observed detection behavior depends strongly on this specific definition, Fig. \ref{fig:uq_detection_90} repeats the analysis using a less extreme threshold, defining catastrophic errors as those exceeding the 90th percentile (LAE$_{90}$).

The resulting recall curves closely resemble those obtained for LAE$_{99}$: fully connected ensembles consistently achieve high recall at low fractions of flagged predictions, whereas latent-evolution ensembles require substantially higher flagged fractions to reach comparable recall.
This demonstrates that the qualitative conclusions regarding uncertainty-based detection efficiency are robust to the precise choice of catastrophic-error threshold.

\section{Iterative evaluation}
\label{app:iterative_eval}

\begin{figure*}
  \centering
  \includegraphics[width=0.9\textwidth]{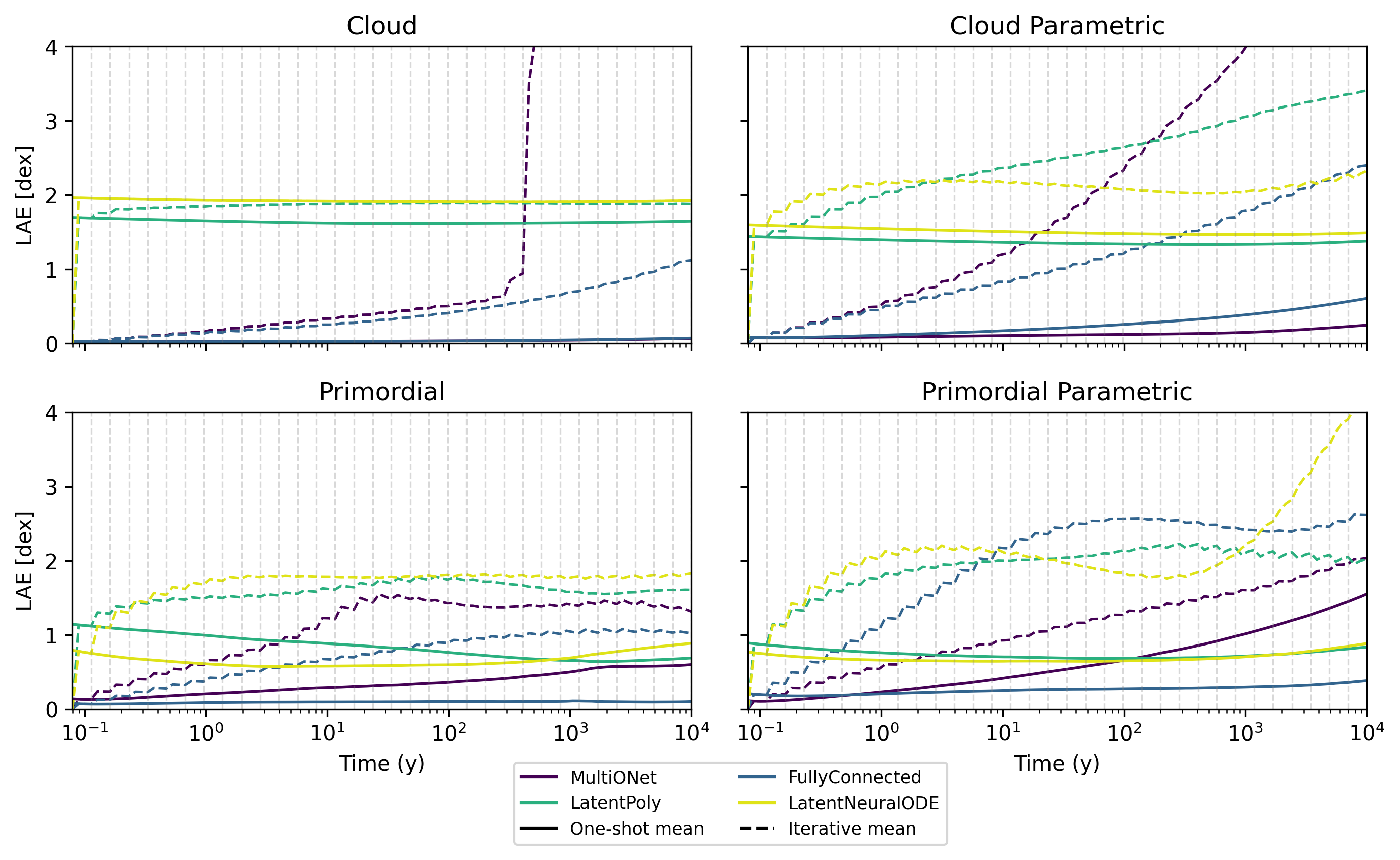}
\caption{Same as Fig. \ref{fig:iterative_errors} but for an interval length of $i=3$ time steps, illustrating stronger sensitivity of error accumulation to more frequent iterative updates.}
  \label{fig:iterative_errors_interval_3}
\end{figure*}

In Sect. \ref{subsec:error_prop} we investigate error accumulation under iterative surrogate application using an interval length of $i=10$ time steps. 
To assess the sensitivity of these findings to the choice of interval length, Fig. \ref{fig:iterative_errors_interval_3} shows the same analysis performed with a shorter interval of $i=3$ time steps, corresponding to more frequent reuse of surrogate predictions as initial conditions.

As expected, reducing the interval length leads to stronger error accumulation across all architectures, reflecting the increased exposure to prediction drift under more frequent iterative updates. 
Importantly, however, the qualitative trends observed in the main text remain unchanged: fully connected surrogates exhibit a more pronounced growth of error over time in relative terms, while latent-evolution architectures retain comparatively stable performance despite their higher initial errors. 
This confirms that the observed differences in iterative robustness are not specific to a particular choice of interval length, but reflect systematic architectural properties.

\end{appendix}
\end{document}